\documentclass[a4paper]{jpconf} 
\usepackage{graphicx} 
\usepackage{bm}
\usepackage{amsmath}
\usepackage{multirow}
\usepackage{subfigure}
\usepackage[normalem]{ulem} 
\usepackage[usenames]{color}
\usepackage{verbatim}

\usepackage[rightcaption]{sidecap}
\graphicspath{ {images/} }

\usepackage[mathscr]{euscript}
\usepackage{mathrsfs}

\usepackage[usenames]{color}
\usepackage{amssymb,amsmath}
\usepackage{tikz}
\usepackage{ifthen}
\usepackage{hyperref}
\hypersetup{
colorlinks=true,
linkcolor=black,
citecolor=blue}

\newcommand{\ud}{\mathrm{d}}
\newcommand{\s}{\sigma}
\newcommand{\be}{\begin{equation}}
\newcommand{\ee}{\end{equation}}

\newcommand{\p}{\prime}
\newcommand{\ii}{\text{i}}

\newtheorem{mytheor}{Theorem}

\newtheorem{mycoroll}{Corollary}

\begin{document}

\title{Scaling theory for anomalous semiclassical quantum transport}

\author{M I Sena-Junior$^{1,2}$ and A M S Mac\^edo$^2$}

\address{ 
$^{1}$Instituto de F\'{i}sica, Universidade Federal Fluminense, 24210-346, Niter\'{o}i, Rio de Janeiro, Brazil\\
$^{2}$ Departamento de F\'{\i}sica, Laborat\'{o}rio de F\'{\i}sica Te\'{o}rica
e Computacional, Universidade Federal de Pernambuco, 50670-901, Recife, Pernambuco,
Brazil}

\ead{marcone@mail.if.uff.br, amsmacedo@df.ufpe.br}

\date{December 2013}

\begin{abstract}
Quantum transport through devices coupled to electron reservoirs can be described in terms of the full counting statistics (FCS) of charge transfer. Transport observables, such as conductance and shot-noise power are just cumulants of FCS and can be obtained from the sample's average density of transmission eigenvalues, which in turn can be obtained from a finite element representation of the saddle-point equation of the Keldysh (or supersymmetric) non-linear sigma-model, known as quantum circuit theory. Normal universal metallic behavior in the semiclassical regime is controlled by the presence of a Fabry-Perot singularity in the average density of transmission eigenvalues. We present general conditions for the suppression of Fabry-Perot modes in the semiclassical regime in a sample of arbitrary shape, a disordered conductor or a network of ballistic quantum dots, which leads to an anomalous metallic phase. Through a double-scaling limit, we derive a scaling equation for anomalous metallic transport, in the form of a nonlinear differential equation, which generalizes the ballistic-diffusive scaling equation of a normal metal. The two-parameter stationary solution of our scaling equation generalizes Dorokhov's universal single-parameter distribution of transmission eigenvalues. We provide a simple interpretation of the stationary solution using a thermodynamic analogy with a spin-glass system.  As an application, we consider a system formed by a diffusive wire coupled via a barrier to normal-superconductor (NS) reservoirs. We observe anomalous reflectionless tunneling, when all perfectly transmitting channels are suppressed, which cannot be explained by the usual mechanism of disorder-induced opening of tunneling channels. 
\end{abstract}


\section{Introduction}
The notion of scaling and its connection to universal behaviour are ubiquitous in physics. In the theory of second order phase transitions the scaling hypothesis, which states that close to the critical point the singular part of the relevant thermodynamic potential is a generalized homogeneous function, predicts universal relations between various critical exponents and a definite procedure for data collapse. Both predictions have met with enormous success in a large variety of systems, both through experimental works and numerical simulations \cite{Stanley1987}. The importance of these results and their contribution to the understanding of phase transitions cannot be overestimated. In the theory of quantum transport in disordered conductors, the scaling hypothesis states that the distributions of transport observables of a sample of size $L$ is fully determined by a single parameter, which in turn can be related to the classical conductance of the sample \cite{Anderson1979, Anderson1980}. In the Landauer-B\"uttiker scattering approach for a two-terminal setup, the single parameter scaling (SPS) hypothesis translates into a single parameter joint distribution of transmission eigenvalues, defined as the eigenvalues $\tau_n$ of the product $tt^{\dagger}$, where $t$ is the sample's transmission matrix \cite{Beenakker1997}. A remarkable prediction of this type of SPS is Dorokhov's universal distribution \cite{Dorokhov1984} for the average density of transmission eigenvalues of a diffusive sample
\be 
\rho(\tau)=\sum_{i=1}^{N_{\text{ch}}}\langle \delta(\tau-\tau_i)\rangle=\frac{g_{cl}}{2 \tau\sqrt{1-\tau}\,},
\label{Dorokhov_density}
\ee 
where $g_{cl}$ is the sample's classical conductance and $N_{\text{ch}}$ is the number of open scattering channels. The bimodality of Dorokhov's distribution and its independence on the sample's shape offers, for instance, the simplest explanation for the universal one-third suppression of shot-noise in phase-coherent diffusive conductors \cite{Beenakker1992}. A striking recent discovery is the observation of the validity of Dorokhov's distribution in ballistic graphene samples \cite{Schuessler2010, Beenakker2006}. 
\par 
Equation (\ref{Dorokhov_density}) can be obtained as the leading term of a semiclassical expansion of the one-point function of transmission eigenvalues, i.e. the universal single variable marginal distribution.  A simple and general way to perform this calculation for a disordered diffusive phase-coherent conductor of arbitrary shape was put forward in \cite{Nazarov1994} using multicomponent Green's functions. In \cite{Macedo2002} a full scaling theory for the crossover from ballistic to diffusive transport in phase-coherent conductors of arbitrary shape was presented, generalizing previous quasi-1D results \cite{Beenakker1994}. The scaling equation reads
\be 
\frac{\partial}{\partial r}\mu(z,r)=-2\frac{\partial}{\partial z}z(1+z)\mu(z,r)f(z,r),\label{scaling}
\ee 
where $\mu(z,r)$ is the average density of the variable $z\equiv -1+1/\tau$ and $r=1/g$ is the dimensionless classical resistance of the sample, furthermore
\be 
f(z,r)=\int_0^{\infty}\ud z^{\prime}\,\frac{\,\mu(z^{\prime},r)\,}{z-z^{\prime}},
\ee 
where the integral is performed as a Cauchy principal value. Applications of this equation include quantum wires and normal-metal-superconductor (NS) junctions with several types of obstacles \cite{Macedo2002, Beenakker1997}. Note that Dorokhov's distribution is a stationary, or fixed point, solution of the scaling equation (\ref{scaling}). Another interesting advance was the description of the ballistic-diffusive crossover using a chain of quantum dots \cite{Belzig2006, Duarte2013}. It was found that the convergence to the continuum behaviour is quite rapid, i.e. a chain with $L \geq 7$ can be described as a diffusive wire, thus exhibiting Dorokhov's distribution. Other recent applications of the transmission eigenvalue density include random media with surface reflections \cite{Xiaojun2013} and diffusive-ballistic crossover for light in absorbing random media \cite{Seng2014}.

\par 
More recently \cite{Sena-Junior2014}, it was found that the continuum limit of a network of quantum dots is more subtle than previously thought. The emergence of Dorokhov's distribution is conditioned to the preservation of Fabry-Perot (FP) modes in the system, i.e. a singular contribution to $\rho(\tau)$ at $\tau=1$. A procedure was proposed to suppress FP modes as we increase the number of dots in the network, similar to the double-scaling limit used to obtain continuum field theories from matrix models \cite{Ginsparg1991}. The final result is a two-parameter family of universal distributions that replaces Dorokhov's density as the leading semiclassical one-point function of transmission eigenvalues of an anomalous diffusive conductor. 
\par
Clearly, equation (\ref{scaling}) is no longer valid in the double-scaling limit. The derivation of a two-parameter scaling equation for the ballistic-anomalous-diffusive crossover under very general conditions is the main objective of this paper. The stationary solution of the two-parameter scaling equation generalizes Dorokhov's universal distribution of transmission eigenvalues for disordered metals and networks of ballistic quantum dots.  The paper is organized as follows. In section 2, we introduce our main results via three general theorems. In Theorem 1 we give an explicit expression for the parameter region which contains a finite density of FP modes. The double-scaling limit, in which the FP modes are suppressed as one approaches the continuum limit, is presented in Theorem 2. Finally, in Theorem 3 we present the scaling equation for the anomalous conductor. We conclude the section by presenting a thermodynamic phase-transition analogy to the stationary solution of the scaling equation. In section 3, we apply the scaling equation to a system formed by a diffusive wire coupled via a barrier to normal-metal-superconductor (NS) reservoirs. We observe anomalous reflectionless tunneling, when all perfectly transmitting channels are suppressed, which cannot be explained by the mechanism of disorder-induced opening of tunneling channels proposed by Y. Nazarov \cite{Nazarov1994}. A summary and conclusions are presented in section 4.

\section{Scaling Theory of Semiclassical Transport}
\par 
In the semiclassical regime, transport properties of a diffusive conductor can be fully described from the density of transmission eigenvalues $\rho(\tau)$. Interestingly, the same description emerges from the continuum limit of a network of ballistic quantum dots \cite{Sena-Junior2014}. It was shown that a necessary condition for the emergence of this universal behaviour is the existence of a singularity in $\rho(\tau)$ at $\tau=1$, which can be related to the sustaining of Fabry-Perot (FP) modes in the system \cite{Macedo2005}. In this section we provide some theorems that lead to a novel kind of scaling equation to describe systems in which the FP modes are suppressed.

We begin by following Y. Nazarov\cite{Nazarov2009} and introducing a complex-valued function, which is related to $\rho(\tau)$, in the regime of large-$N_{\text{ch}}$, as 
\begin{equation}\label{spectral_current}
I(\varphi)=\int_{0}^{1}\ud\tau\,\rho(\tau)\,\frac{\tau\,\sin(\varphi)}{1-\tau\sin^2(\varphi/2)}.
\end{equation}
In circuit theory $I(\varphi)$ is interpreted as a pseudo-current which satisfies Kirchhoff type of conservation laws. In order to keep track of the singularity at $\tau=1$, we parametrize the transmission eigenvalues through $\tau=\text{sech}^2 x$, thus the new density $\nu(x)$ is related to $\rho(\tau)$ via
\begin{equation}
\rho(\tau)=\frac{\nu(x)}{2\tau\sqrt{1-\tau}}\Biggl\vert_{\text{sech}^2 x=\tau}.
\end{equation}
Note that if $\rho(\tau)$ exhibits an inverse square-root singularity at $\tau=1$ (or $x=0$), than $\nu(0)$ is finite. The value of $\nu(0)$ can thus be interpreted as a density of FP modes. We may obtain $\nu(x)$ directly from the pseudo-current $I(\varphi)$ using the formula
\begin{equation}\label{nu}
\nu(x)=\frac{1}{\pi}\,\text{Re} \Bigl[ I(-2\ii x+\pi-0^{+})\Bigl].
\end{equation}
Our first theorem establishes the support of $\nu(0)$ in an arbitrarily shaped conductor connected to two electron reservoirs via barriers of arbitrary transparencies. 

\begin{mytheor}[Fabry-Perot modes]\label{theor_1}
Consider a two-terminal setup consisting of a sample (a disordered conductor or a network of quantum dots), with dimensionless conductance $g_{S}$, connected to normal reservoirs, left (L) and right (R), via two connectors of dimensionless conductance $g_{\s} \equiv N_{\s}\, T_{\s}$,  with $\s=L, R$. The parameters $N_{\s}$ and $T_{\s}$ are the number of transmission channels and the transparency of the $\s$-connector, respectively. Pseudo-current conservation yields
\begin{equation}\label{kirchhoff}
I(\phi)\equiv I(\phi-\theta; T_{L})=I_{S}(\theta-\theta^{\prime})=I(\theta^{\prime}; T_{R}),
\end{equation} 
where $I(\varphi ; T_{\s})$ is the pseudo-current of the $\s$-connector, defined by equation \eqref{spectral_current} with  $\rho(\tau;T_{\s})=N_{\s}\,\delta(\tau-T_{\s})$. Furthermore, $I_{S}(\theta-\theta^{\p})$ is the pseudo-current of the sample defined by equation \eqref{spectral_current} with an arbitrary density of transmission eigenvalues, $\rho_{S}(\tau)$. Then, the support of $\nu(0)$, i.e. the parameter region where $\nu(0)>0$, is given by $0\leqslant\zeta<\zeta_{c}$, where
\begin{align}\label{zeta_LR}
&\frac{\zeta}{\,N_{\text{ch}}^{(\ast)}\,}\equiv\left\vert\frac{1}{g_L}-\frac{1}{g_R}\right\vert& &\text{and}& &\zeta_{c} \equiv 1 + \frac{1}{g_{S}/N_{\text{ch}}^{(\ast)}}\, , &
\end{align}
where $N_{\text{ch}}^{(\ast)}\equiv N_{R}\,\Theta(g_{L}-g_{R}) + N_{L}\,\Theta(g_{R}-g_{L})$.
\end{mytheor}

A proof is presented in \ref{Appen_A1}. Particular cases can be found in previous works, for instance, a numerical verification for a linear chain of ballistic quantum dots is shown in \cite{Duarte2013} and an analytical study for a network of connected chains of ballistic quantum dots can be found in \cite{Sena-Junior2014}. 

\par
In order to gain some understanding on how Theorem \ref{theor_1}  works, let us analyse in detail the simple case of a ballistic quantum dot with inequivalent asymmetric barriers ($T_{L}\neq T_ {R}$, $N_{L}\neq N_ {R}$).  Then, equation \eqref{kirchhoff} reads $I(\phi)\equiv I_{L}(\xi)=I_{R}(\xi^{\p})$ for  $\phi=\pi-0^{+}$ and $\xi\equiv\xi^{\p}$, where
\begin{subequations}\label{I_LR}
\begin{align}
I_{L}(\xi)&\equiv\frac{2\,g_{L}\,\xi}{1-T_{L}+\xi^2}=I(\pi-\theta;T_L)\quad\,\text{with}\quad\xi=\tan(\theta/2),&\\
I_{R}(\xi^{\p})&\equiv\frac{2\, g_{R}\,\xi^{\p}}{1+(1-T_{R})\,\xi^{\p\,2}}=I(\theta^{\p};T_{R})\quad\text{with}\quad\xi^{\p}=\tan(\theta^{\p}/2),&
\end{align}
\end{subequations}
\noindent in which $g_{\s}=N_{\s}T_{\s}$ is the conductance of the $\s-$connector and $\theta\equiv\theta^{\p}$ is the pseudo-potential in the quantum dot. Once  $\xi$ is determined from the equation $I_{L}(\xi)=I_{R}(\xi)$, the density of FP modes, $\nu(0)$, is calculated from \eqref{nu} yielding
\begin{equation}\label{FP_1QD}
\nu(0)=\frac{1}{\pi}\text{Re}\left[I(\pi-0^{+})\right]=\frac{2}{\pi}\,\frac{\sqrt{N_{L} N_{R}}}{1/T_{L}+1/T_{R}-1}\,\sqrt{(1-\zeta_{L})(1-\zeta_{R})\;}\,\Theta(1-\zeta_{L})\Theta(1-\zeta_{R}),
\end{equation}
where we defined the auxiliary variables $\zeta_{\s}\equiv N_{\s}\left(1/g_{\s}-1/g_{\overline{\s}}\right)$ with $\lbrace\s,\overline{\s}\rbrace = L,R$. For equivalent barriers, we have $N_{L}=N_{R}$ and equation \eqref{FP_1QD} agrees with equation (86) in \cite{Macedo2005}. We have thus explicitly found that FP modes are present only in the region $0\leqslant\zeta<1$, where $\zeta\equiv\zeta_{R}\,\Theta(g_L - g_R)+\zeta_{L}\,\Theta(g_R - g_L)$ in agreement with equation \eqref{zeta_LR}, since $\zeta_{c}=1$ for a ballistic quantum dot. A ballistic quantum dot is accommodated in Theorem \ref{theor_1} through the limit $\theta\to\theta^{\p}$ in \eqref{kirchhoff}. This condition means that there is no pseudo-potential drop inside the quantum dot, which is consistent with circuit theory if we take $g_{S}\to\infty$, which according to equation \eqref{zeta_LR} implies $\zeta_{c}\to 1$, as stated before. The boundary of the region with FP modes ($\nu(0)=0^{+}$) is given by $\zeta=\zeta_{c}-0^{+}$.

Interestingly, the density of FP modes shown in \eqref{FP_1QD} exhibits a power law behavior $\nu(0)\propto (1-\zeta)^{1/2}$ for $\zeta=1 - 0^{+}$ and a full suppression, $\nu(0)=0$, for $\zeta>1$. The value of the exponent will be discussed later with a thermodynamic analogy. We present below some corollaries to Theorem 1 with several types of power law behaviors.

\begin{mycoroll}[Power law I]\label{coro_1}
The system described in Theorem \ref{theor_1} displays a density of FP modes with a power law of the form $\nu(0)\propto (\zeta_c -\zeta)^{1/2}$ for $\zeta=\zeta_c - 0^{+}$ and $\nu(0)=0$ for $\zeta>\zeta_c$. 
\end{mycoroll}

A proof is shown in \ref{Appen_A2}. As pointed out in \cite{Macedo2005}, we can interpret the emergence of FP modes in the sample as a kind of second order phase transition, with $\zeta$ playing the role of temperature and $\zeta_{c}$ being analogous to a critical temperature. The density of the FP modes, $\nu(0)$, behaves as an order parameter, being zero for $\zeta>\zeta_{c}$ and taking on a finite value for $\zeta<\zeta_{c}$. Then, the power law behaviour $\nu(0)\propto (\zeta_{c}-\zeta)^{\beta}=0^{+}$, with $\beta=1/2$ is consistent with a mean field description of the corresponding phase transition.

\begin{mycoroll}[Power law II]\label{coro_2} 
The system described in the Theorem \ref{theor_1} has an average density of the variable $x\equiv{\rm arcsech}\sqrt{\tau}=0^{+}$, where $\tau$ is a transmission eigenvalue, that satisfies a power law  $\nu(x)\propto x^{1/3}$ at the line $\zeta=\zeta_c$.
\end{mycoroll}

The proof is given in \ref{Appen_A3}. Thus, the average density $\nu(x)$ behaves as $\nu(x)\propto x^{1/\delta}$, in analogy with magnetic system close to criticality, where the magnetization $M$ shows a power law behavior as a function of the external field $H$ in the critical isotherm  $M\propto H^{1/\delta}$. The mean-field value of the exponent is $\delta=3$.

\par
\begin{mycoroll}[Power law III]\label{coro_3}
The system described in Theorem \ref{theor_1} exhibits the power law behavior $x\propto (\zeta-\zeta_{c})^{3/2}=0^{+}$ for $\nu(x)=0^{+}$.
\end{mycoroll}

The proof is given in \ref{Appen_A6}. Note that for $\zeta=\zeta_{c} + 0^{+}$ and $x=0^{+}$ the power law $\nu(0^{+})\propto (\zeta-\zeta_{c})^{\beta^{\p}}=0^{+}$ is valid for $\beta^{\p}=1/2$, which agrees with the scaling prediction $\beta^{\p}=\beta$. From a simple scaling arguments, we find $\nu(x)\propto(\zeta - \zeta_{c})^{\beta}$ and $\nu(x)\propto x^{1/\delta}$ at $\zeta=\zeta_{c}$, so that $x$ scales as $(\zeta-\zeta_{c})^{\Phi/2}\,$ for $\,\Phi/2 = \beta\,\delta = 3/2$. Note that this latter relation is similar to the scaling law of spin glasses with $\Phi$ being the crossover exponent  \cite{Fisher}. The mean-field value $\Phi=3$ is the known result for the Almeida-Thouless line \cite{Almeida1978}.  

We are now in position to introduce the double-scaling limit that leads to anomalous metallic behavior in arbitrarily shaped samples.

\begin{mytheor}[double-scaling Limit]\label{theor_2}
Consider the system described in Theorem \ref{theor_1}. Let $s$ be the length of the sample in units of the mean free path and $N_S$ be the number of transmission channels. The double-scaling limit is defined by taking the limits $s \to \infty$ and $N_S \to \infty$ keeping fixed the sample's conductance $g_S\equiv N_{S}/s$ and the parameter $\lambda\equiv\zeta_{c}/\zeta$. Then, the pseudo-current of the sample is given by
\begin{equation}\label{diffusive}
I(\phi)=g_{S} \left(\phi-\bar{\phi}_{\lambda}\right),
\end{equation}
where $\bar{\phi}_{\lambda}=\bar{\phi}\,\Theta(1-\lambda)+0^{+}$, $\phi=\bar{\phi}+\lambda\,\sin\bar{\phi}\,$,  $\Theta(x)$ is the Heaviside function and we set $\,\Theta(0)\equiv 1$. 
\end{mytheor}

The proof is given in \ref{Appen_A3}. The variable $\lambda$ plays the role of a suppression parameter of the FP modes and the variable $s$ is the length scale parameter of the sample. For the condition $\lambda>1$, we obtain a linear relationship $I(\phi)$ in \eqref{diffusive}, which is known to be valid for the usual diffusive regime. This regime is obtained through the limit ($s\gg 1$) through a finite density FP modes, ie $\nu(0)>0$, and according to Theorem \ref{theor_1}. On the other hand, for the condition $0<\lambda\leqslant 1$, we find a non-linear relationship $I(\phi)$, which brings up an anomalous diffusive regime, for which the diffusive limit is accompanied by the supression of FP modes, i.e. $\nu(0)=0$. In \cite{Sena-Junior2014}, equation \eqref{diffusive} is explicitly obtained for the particular topology of coupled chains of ballistic quantum dots. Here we present arguments for the validity of these results for an arbitrary sample. We show explicitly the extended universality established in the diffusive regime of an arbitrary quantum network. The distribution of transmission eigenvalues is sensitive to the change of universality labeled by suppression parameter $\lambda$, as discussed below.

\begin{mytheor}[Scaling equation]\label{scale}
Consider a two-terminal system that consists of a sample, or simply barriers, in series with an anomalous diffusive conductor. Let $I_{0}(\phi)$ be the pseudo-current of an auxiliary sample obtained by neglecting the potential drops across all sections of the anomalous diffusive conductor. Then, the pseudo-current of the original system satisfies
\begin{equation}\label{sol_implicit}
I(\phi;r)=I_{0}\left(\phi-\bar{\theta}_{\lambda} - r I(\phi; r)\right),\quad\text{where}\quad\sin\bar{\theta}_{\lambda}=\frac{r\, I(\phi;r)}{\lambda}\,\Theta(1-\lambda)+0^{+},
\end{equation}
which is an implicit solution of the following differential equation

\begin{equation}\label{scale_dif}
\frac{\partial I(\phi;r)}{\partial r} + \left(1+\frac{\Theta(1-\lambda)}{\sqrt{\lambda^2 - r^2 I^2(\phi;r)}}\right)I(\phi;r)\frac{\partial I(\phi;r)}{\partial\phi}=0,
\end{equation}
with initial condition $I(\phi,0)=I_{0}(\phi)$.
\end{mytheor}

Equation \eqref{scale_dif} is the main result of this paper. It is the semiclassical scaling equation that describes the evolution of the pseudo-current as a function of the dimensionless classical resistance of the anomalous metal. We proceed by implementing the map $z=-\cos^2(\phi/2)$ and defining $G(z,r)=-I(\phi,r)/\sin\phi$, so that we can rewrite equation \eqref{scale_dif} as
\begin{equation}\label{equation_G}
\frac{\partial G(z,r)}{\partial r} + \frac{\partial}{\partial z}\left[z(1+z)\,G^{2}(z,r)\,+\,\frac{\Theta(1-\lambda)}{r}\,\sqrt{\frac{\lambda^2}{4\,r^2} + z(1+z)\,G^{2}(z,r)\,}\,\right]=0.
\end{equation}
We can now obtain the scaling equation for the density of transmission eigenvalues, $\rho(\tau,r)$, through the identity $G(1/\tau-1 \pm\ii\,0^{+},r)=V(\tau,r)\,\mp\,\ii\,\pi\, \rho(\tau,r)$. Note that for $\lambda>1$, i.e. when FP modes are not suppressed,  \eqref{equation_G} yields \eqref{scaling}, as expected \cite{Macedo2002}. However, for $0<\lambda\leqslant 1$, the last term of  \eqref{equation_G} accounts for the behavior of the anomalous metal.

We can gain further intuition on the physical meaning of \eqref{scale_dif} by introducing the following change of variables: $J(\phi;u)\equiv r\,I(\phi,r)$ with $u\equiv \ln r$. Then, \eqref{scale_dif} becomes the following equation for a one-dimensional hydrodynamic flow
\begin{equation}\label{bacteria}
\left[\,\frac{\partial}{\partial u}+\beta(J)\,\frac{\partial}{\partial\phi}-1\,\right]\,J(\phi;u)=0,
\end{equation}
where $\beta\left(J\right)=J+\dfrac{J}{\sqrt{\lambda^{2}-J^{2}}}\,\Theta(1-\lambda)$, with the shorthand notation $J=J(\phi,u)$. The diffusive regime corresponds to the asymptotic solution, which is obtained by setting $\partial_{u}J(\phi;u)=0$ in \eqref{bacteria} which ends up giving equation \eqref{diffusive}. Note that the asymptotic solution is linear in the pseudo-potential, $J(\phi)=\phi$, for $\lambda>1$, and non-linear for $0<\lambda\leqslant 1$ with the constraint $0\leqslant J(\phi)<\lambda$.

\subsection*{Density of transmission eigenvalues}

\begin{figure}[h!]\centering
\includegraphics[width=0.50\textwidth]{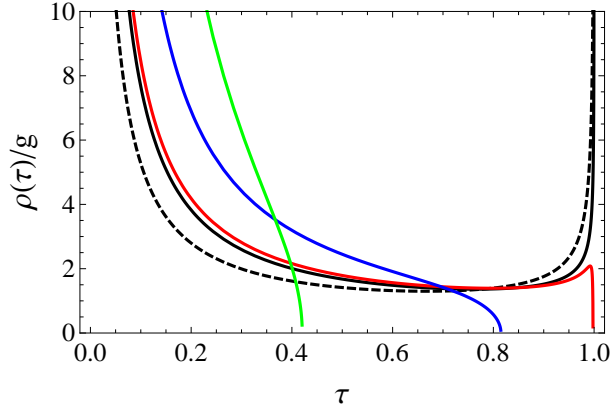}
\caption{Distributions of transmission eigenvalues in the anomalous metal regime. The distribution $\rho(\tau)$ is normalized by conductance $g=g_{cl}\,\left[1-(1+\lambda)^{-1}\,\Theta(1-\lambda)\right]$, where $g_{cl}$ is the classical conductance of usual metallic regime ($\lambda>1$). The dashed curve corresponds to Dorokhov's distribution. The curves in black, red, blue and green correspond to to $\lambda=1.0$, $\lambda=0.8$, $\lambda=0.3$ and $\lambda=0.1$, respectively (color online).} \label{fig:distrib_rho}
\end{figure}

In the usual metallic regime ($\lambda>1$), we find $\nu(x)=g_{cl}$, where $g_{cl}$ is the system's classical conductance, from which we obtain Dorokhov's distribution $\rho(\tau)$, shown in \eqref{Dorokhov_density}.
For the anomalous metallic regime ($0<\lambda\leqslant 1$), we write the density as $ \nu (x) = g_{cl}\,\tilde{\nu}(x)$, and use the parametrization $\phi = -2 \ii\,x + \pi$ and $\bar{\phi}\equiv\pi(1- \tilde{\nu}) + \ii\,b$ in the relation $\phi = \bar{\phi} + \lambda \sin\bar{\phi}$ to obtain the equations $\lambda\sin\pi\tilde{\nu}\,\cosh b=\pi\tilde{\nu}$ and $b -\lambda\cos\pi\tilde{\nu}\,\sinh b=-2x$. Eliminating the variable $b<0$, we obtain for $\tilde{\nu}(x)$ the following implicit equation
\begin{equation}\label{implicit}
\text{arccosh}\left(\frac{\pi\tilde{\nu}(x)}{\lambda\sin\pi\tilde{\nu}(x)}\right) - \lambda\cos\pi\tilde{\nu}(x)\,\sqrt{\left(\frac{\pi\tilde{\nu}(x)}{\lambda\sin\pi\tilde{\nu}(x)}\right)^{2}-1} = 2x.
\end{equation}
The corresponding density of transmission eigenvalues, $\rho(\tau)$,  for different values of the $\lambda$ are shown in figure \ref{fig:distrib_rho}. 

Note that the distributions present a bimodal behavior for $\lambda\geqslant 1$ with peaks at $\tau=0$ and $\tau=1$. Near the second peak, we can show that $\rho(\tau)\propto (1-\tau)^{-1/3}$ for $\lambda=1$, while $\rho(\tau)\propto (1-\tau)^{-1/2}$ for $\lambda>1$ (Dorokhov's distribution). The domain region that defines the suppression, $\rho(\tau)=0$ for $0<\lambda<1$, is given by $\tau^{\ast}<\tau\leqslant 1$, where $\tau^{\ast}\equiv\text{sech}^{2}x^{\ast}$ and $\tilde{\nu}(x^{\ast})=0^{+}$, which can be obtained from \eqref{implicit}.

The set of distributions for $0<\lambda\leqslant1$ is a novel two-parameter universality class for the transmission eigenvalue density, for which the suppression of perfect transmission channels (or Fabry-Perot modes) plays a key role. In order to gain more understanding into the meaning of the anomalous metallic regime, we present in the next section an insightful analogy with the thermodynamics of a spin-glass system.

\subsection*{Phase Transition analogy}

\begin{figure}[h!]\centering
\includegraphics[scale=0.50]{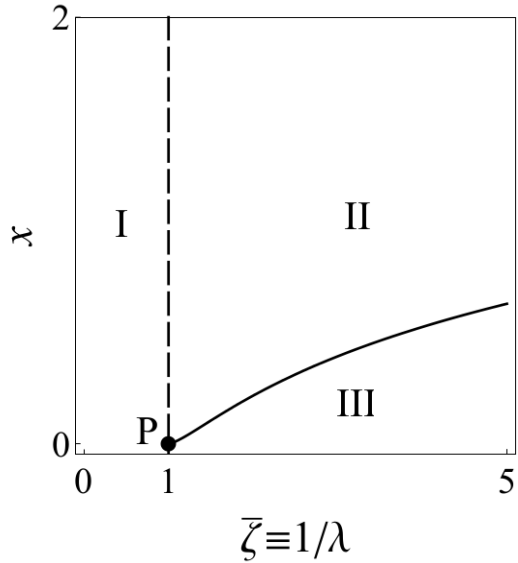}
\caption{Regions I, II and III correspond to an ordered phase ($\tilde{\nu}=1$), a partially ordered phase ($0<\tilde{\nu}<1$) and a disordered phase ($\tilde{\nu}=0$) respectively. The dashed line at $\bar{\zeta}=1$  is a first order transition line. The solid line between the phases II and III corresponds to a second order transition line. The intersection of the two curves occurs at a tricritical point P.} \label{fig:network}
\end{figure}

In analogy to phase transitions in thermodynamics, equation \eqref{implicit} can be interpreted as an ``equation of state'' that relates the ``order parameter'' $\tilde{\nu}$, the  ``external field'' $x$ and the ``effective transition parameter (temperature)'' $\bar{\zeta}\equiv 1/\lambda\geqslant 1$. In figure \ref{fig:network} we show a ``phase diagram'' with regions characterizing three different  behaviors of $\tilde{\nu}$ in the intervals $x\geqslant 0$ and $\bar{\zeta}> 0$. 

We can interpret the diagram as follows: region I ($0<\bar{\zeta}<1$) corresponds to an ordered phase, where the order parameter is frozen ($\tilde{\nu}=1$) ; region II ($\bar{\zeta}\geqslant 1$ and $x>x^{\ast}$) is a partially disordered phase, in which there is a competition between the effects of the ``external field'' $x$ and the ``effective temperature'' $\bar{\zeta}$ on the ``order parameter'' with values in the interval $0<\tilde{\nu}<1$; region III ($\bar{\zeta}\geqslant 1$ and $x\leqslant x^{\ast}$) corresponds to a disordered phase for which $\tilde{\nu}\equiv 0$. Note that in region II a saturation of the order parameter ($\tilde{\nu}\to 1$) occurs for $x\gg 1$. The separatrix curve between phases II and III is defined by $\nu(x^{\ast})=0^{+}$ for  $\bar{\zeta}\geqslant 1$ in \eqref{implicit}. We get 
\begin{equation}\label{fronteira_FP}
x^{\ast}(\bar{\zeta})=\frac{1}{2}\left[\,\text{arccosh}\left(\bar{\zeta}\right)-\sqrt{1-\frac{1}{\bar{\zeta}^2}}\,\right].
\end{equation} 
The ``isothermal behavior'' ($\bar{\zeta}=\text{constant}$) for $\nu(x)$ near the separatrix  $x^{\ast}(\bar{\zeta})$ can be described with the following power laws:
\begin{equation}
\tilde{\nu}(x)=\begin{cases}
D_{c}\,x^{1/\delta},\qquad\quad\;\,\bar{\zeta}=1\;\text{and}\;x = 0^{+};\\
D_{+}\,(x-x^{\ast})^{\beta},\;\;\;\bar{\zeta}>1\;\text{and}\;x-x^{\ast} = 0^{+},\end{cases}
\end{equation}
with mean-field values for the exponents $\delta=3$ and $\beta=1/2$. In the ``isochore'' ($x=\text{constant}$), we find $\tilde{\nu}=B\,(-t)^{\beta}$ (where $t\equiv\bar{\zeta}-\bar{\zeta}^{\ast}$). The non-universal amplitudes take on the values: $D_{c}=\frac{3^{5/6}}{2^{1/3}\pi}\simeq 0.631114$, $D_{+}=\frac{2}{\pi \left(1-\lambda^2\right)^{1/4}}$ and $B=\frac{\sqrt{2 \lambda}}{\pi}$.

The separatrix curve between phases II and III is analogous to the Almeida-Thouless line \cite{Almeida1978, DeDominicis2006}, which appear in phase diagrams of spin glasses (SG) and separates the ordered phase from the disordered one in the presence of an external field. 
In the neighborhood of the tricritical point P ($x=0$ and $\bar{\zeta}=1$) represented in the $x-\bar{\zeta}$ diagram, we obtain a power law
\begin{equation}
x^{\ast}\simeq\frac{\sqrt{2}}{3}\,\left(\bar{\zeta}-1\right)^{\,\Phi/2}\quad\text{for}\quad\bar{\zeta}\to 1^{+},
\end{equation}  
where $\Phi=3$ coincides with the corresponding mean field exponent of a spin glass, in which case the power law reads  $H\propto (T-T_{c})^{\Phi/2} $, where $H$ is an external magnetic field and $T$ is the temperature of the system.

We conclude this section by remarking that this phase-transition analogy with mean-field exponents is a surprising feature of the extended two-parameter universality class of transmission eigenvalue densities. In the next section we show how to use the scaling law to describe anomalous-metal-superconductor systems.

\section{Anomalous-metal-superconductor systems}

The full counting statistics (FCS) of charge transfer through a normal-superconductor (NS) device at energies much lower than the Thouless energy can be described, in the absence of an external magnetic field, in terms of the generating function $\Phi_{\text{NS}}(\eta)=\sum_{i=1}^{N_{\text{ch}}}\ln\left[1+R_{j}\left(\text{e}^{\ii \eta} - 1\right)\right]$, where $\lbrace R_{j}\rbrace$ are Andreev reflection eigenvalues. The average FCS cumulants,  $\lbrace (q_{n})_{\text{NS}}\rbrace$, can be obtained from
\begin{subequations}
\begin{equation}\label{gen_ns}
S_{\text{NS}}(\eta)\equiv\langle\Phi_{\text{NS}}(\eta)\rangle \equiv \sum_{n=1}^{\infty}\,\frac{(\ii\eta)^n}{n!}\,(q_{n})_{\text{NS}}\quad\therefore\quad (q_{n})_{\text{NS}}=\frac{\partial^{n}}{\ii^{n}\partial\eta^{n}}S_{\text{NS}}(\eta)\Biggl\vert_{\eta=0}.
\end{equation}
It is possible to relate the Andreev eigenvalues $\lbrace{R_{j}\rbrace}$ to the transmission eigenvalues of the normal side $\lbrace{\tau_{j}\rbrace}$ by $R_{j} =\tau_{j}^{2}/\left(2-\tau_{j}\right)^2$, which is very useful for calculating the average cumulants of charge transport in SN systems \cite{Beenakker1997}. A simple consequence of this relation is the connection between the pseudo-current of the NS system to the pseudo-current of the corresponding NN system. We find
\begin{equation}\label{current_ns}
I_{\text{NS}}(\phi)=-2\frac{\partial}{\partial\phi}S_{\text{NS}}(\eta)\Biggl\vert_{\text{e}^{\ii\eta}=\cos^{2}(\phi/2)}=\frac{1}{2}\,\sum_{\s=\pm}I(\phi_{\s}),\quad\text{with}\quad \phi_{\s}=\phi/2+\s\,\pi/2,
\end{equation}
\end{subequations}
where $I(\varphi)$ is the pseudo-current of corresponding system with normal reservoirs. From \eqref{gen_ns} and \eqref{current_ns} we get
\begin{equation}
(q_{n})_{\text{NS}}=\left(\frac{\varepsilon^2 -1}{2\varepsilon}\frac{\ud}{\ud\varepsilon}\right)^{n-1} \frac{\sqrt{1-\varepsilon^{2}}}{2\varepsilon}\,I_{\text{NS}}(\phi)\Biggl\vert_{\varepsilon=0}\quad\text{with}\quad \sin(\phi/2)=\varepsilon.
\end{equation}
We can also introduce an average density of Andreev reflections eigenvalues, $\rho_{\text{NS}}(r)$, which can be calculated from the pseudo-current using the formulas
\begin{equation}
\rho_{\text{NS}}(r)=\frac{\nu_{\text{NS}}(x)}{2r\sqrt{1-r}}\Biggl\vert_{\text{sech}^2x= r},
\end{equation}
where
\begin{equation}\label{nu_NSA}
\nu_{\text{NS}}(x)=\frac{1}{\pi}\text{Re}\left[I_{\text{NS}}(-2\ii x + \pi +0^{+})\right].
\end{equation}
Note that $\nu_{\text{NS}}(0)$ can be interpreted as the density of Fabry-Perot modes in the system.

\subsection*{The diffusive regime}

The diffusive regime is described by the stationary solution of the scaling equation \eqref{scale_dif}, i.e by inserting \eqref{diffusive} in \eqref{current_ns}. The resulting pseudo-current of the diffusive NS-system is
\begin{equation}\label{pseudo_current_difusivo_NS}
I_{\text{NS}}(\phi)=\frac{g}{2}\sum_{\s=\pm}\left(\phi_{\s}-\bar{\phi}_{\lambda,\s}\right),
\end{equation}
where $\phi_{\s}=\left(\phi+\s\,\pi\right)/2$ and $\bar{\phi}_{\lambda,\s}=\bar{\phi}_{\s}\,\Theta(1-\lambda)+0^{+}$ besides the constraint $\phi_{\s}=\bar{\phi}_{\s}+\lambda\,\sin\bar{\phi}_{\s}$. For the usual diffusive regime ($\lambda>1$), the pseudo-current in \eqref{pseudo_current_difusivo_NS} is just half that of a similar system with normal reservoirs and thus, the ratios of charge transfer cumulants, $\lbrace (q_{n+1}/q_{1})_{\text{NS}}\rbrace$, have the same values as those of the corresponding NN system. Inserting \eqref{pseudo_current_difusivo_NS} in \eqref{nu_NSA} we obtain a density of Andreev's reflection eigenvalues given by $\nu_{\text{NS}}(x)=\nu(2x)/2$, where $\nu(x)$ is the density of transmission eigenvalues of the NN system. Thus, we find that the boundary of the FP modes is the same for the NN and NS systems, $x^{\ast}(\bar{\zeta})=0$ or $\bar{\zeta}=1$ (according to equation \eqref{fronteira_FP}).
\par 
Let us now calculate the charge transfer cumulants of the NS system in the anomalous diffusive regime, i.e. with suppression of FP modes ($0<\lambda\leqslant 1$). Note that we can express the pseudocurrent $I_{\text{NS}}(\phi)$ in \eqref{pseudo_current_difusivo_NS} through the variables
$\bar{\phi}_{\s}=\Psi_{+}-\sigma\,\Psi_{-}+\sigma\,\pi/2$ (for $\s=\pm$) as $I_{\text{NS}}(\phi)=g\,(\phi/2-\Psi_{+})$, provided the following transcendental relations are satisfied: $\Psi_{+}+\lambda\,\sin\Psi_{+}\,\sin\Psi_{-}=\phi/2$ and $\Psi_{-}=\lambda\,\cos\Psi_{-}\,\cos\Psi_{+}$. We proceed by representing the variables $\Psi_{+}$ and $\Psi_{-}$ in series of powers of $\varepsilon=\sin(\phi/2)$ so that $\Psi_{+}=\sum_{n\geqslant 1} a_{2n-1}\,\varepsilon^{2n-1}$ and $\Psi_{-}=\Psi^{\ast}+\sum_{n\geqslant 1}b_{2n}\,\varepsilon^{2n}$, where the coefficient $\Psi^{\ast}$ satisfies $\Psi^{\ast}=\lambda\,\cos\Psi^{\ast}$. It is now straightforward to calculate the charge transfer cumulants. We show below the first four cumulants, $\lbrace(q_{n})_{\text{NS}}\rbrace$ as a function of $\lambda$
\begin{subequations}
\begin{align}
\frac{\,(q_1)_{\text{NS}}\,}{g/2}=&\,1-\frac{\,\Theta(t)\,}{z},\label{Q1}\\
\frac{(q_2)_{\text{NS}}}{g/2}=&\,\frac{1}{3}-\frac{\Theta(t)}{z}\,\left(\frac{t}{2\, z^4}-\frac{5}{6\,z^3}+\frac{1}{3\,z^2}+\frac{1}{3}\right),\label{Q2}\\
\frac{(q_3)_{\text{NS}}}{g/2} =&\,\frac{1}{15}-\frac{\Theta(t)}{z}\left(\frac{7 t^2}{4\,z^8}-\frac{21 t}{4\,z^7}+\frac{11+6t}{3\,z^6}-\frac{77}{30\,z^5}+\frac{4+5t}{10 \,z^4}-\frac{5}{6\,z^3}+\frac{1}{3\,z^2}+\frac{1}{15}\right),\label{Q3}\\
\frac{(q_4)_{\text{NS}}}{g/2} =&-\frac{1}{105}\,- \frac{\Theta(t)}{z}\,\Biggl(\frac{99\, t^3}{8\, z^{12}}-\frac{429\,t^2}{8\,z^{11}}+\frac{3t\,(98+27t)}{4\,z^{10}}-\frac{935+1584 t}{30\,z^{9}} -\frac{669+735t}{70\,z^7}\nonumber\\ 
&\qquad\qquad\qquad\qquad  +\frac{4(43+21t)}{21\,z^6}-\frac{77}{15\,z^5}+\frac{8+3t}{10\,z^4}-\frac{1}{2 \,z^3}+\frac{1}{5\,z^2}-\frac{1}{105}\Biggl),\label{Q4}
\end{align}
\end{subequations}
where $t\equiv 1-\lambda^2$, $z\equiv 1+\lambda\sin\Psi^{\ast}$ and we defined $\Theta(0)\equiv1$.

\begin{figure}[]
\center\includegraphics[width=0.50\textwidth]{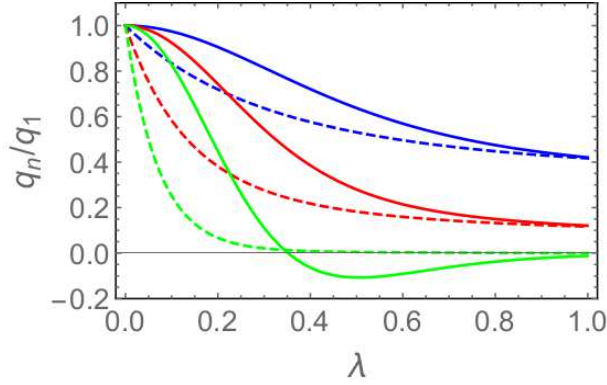}
\caption{Cumulant ratios $q_{2}/q_{1}$, $q_{3}/q_{1}$ and $q_{4}/q_{1}$ as a function of $\lambda$ for $0<\lambda\leqslant 1$. The solid (dashed) lines correspond to the NS (NN) sytem. The blue, red and  green curves corresponds to $n =2,\, 3\,\text{and}\,4$, respectively. Note that at $\lambda = 1$ the NS and NN values are very close (color online).}
\label{fig_cumulantegamma}
\end{figure}

Note that all cumulants have a discontinuity at $\lambda=1$ due to the first-order transition in the density $\nu_{\text{NS}}(x)$, where $\text{sech}^2(x)=R$. For $\lambda>1$ (or $t<0$) the set of cumulant ratios, $(q_{n})_{\text{NS}}/(q_{1})_{\text{NS}}$, reproduce the values of the normal diffusive quantum wire, which are $1/3$, $1/15$, $-1/105$, $\ldots\,$, respectively for $n=2,\,3,\,4,\, \ldots\,$, (see e.g. \cite{Sena-Junior2014}). The values of the cumulant ratios for $0<\lambda\leqslant 1$ (or $0\leqslant t<1$), where FP modes are suppressed, are distinct from those of the corresponding system with NN-reservoirs. In figure \ref{fig_cumulantegamma} we compare the plots of the first three cumulant ratios, for NS and NN systems, as a function of $\lambda$. 

Finally, we may write the cumulants $\lbrace(q_{n})_{\text{NS}}\rbrace$,  explicitly in terms of the variable $\lambda$ through an approximate solution for the auxiliary variable, $\Psi^{\ast}$, given by $\Psi^{\ast}=\lambda\cos\Psi^{\ast}\approx\lambda\left(1-(\Psi^{\ast})^{2}/2\right)$, thus $\Psi^{\ast}\approx\left(\omega-1\right)/\lambda$, where $\omega\equiv\sqrt{1+2\lambda^2}$. Therefore, we can write the auxiliary parameter $z$ as $z\approx 1+ \lambda\left(\Psi^{\ast}-(\Psi^{\ast})^{3}/6\right)=\omega-(\omega-1)^{2}/(3\omega-3)$. Then, from \eqref{Q1}-\eqref{Q4}, we can express the ratios of cumulants $(q_{n})_{\text{NS}}/(q_{1})_{\text{NS}}$ as follows
\begin{equation}\label{q_NS_eq}
\frac{(q_{n})_{\text{NS}}}{(q_{1})_{\text{NS}}}\approx \Biggl(\frac{q_{n}}{q_{1}}\Biggl)_{\substack{
   \text{usual}\;\;\\
   \text{wire}
  }}\,+\,\, \Theta(1-\lambda)\,\frac{(1+\omega)^{3}}{(\omega +2) \left(2 \omega ^2+5 \omega -1\right)^{4(n-1)}}\,P_{n}(\omega),\quad\text{for}\quad n=2, 3, \ldots
\end{equation}
where the first three polinomials $P_{n}(\omega)$ (for $n=$ 2, 3 \text{and} 4) are shown in Appendix B in equations \eqref{P2}, \eqref{P3} and \eqref{P4}. The approximate values of the first three ratios $(q_{n})_{\text{NS}}/(q_{1})_{\text{NS}}$ at $\lambda=1$ ($\,\omega=\sqrt{3}\,$) are $F=\frac{127}{300}=0.423333$, $S=\frac{11407}{93750}=0.121675$ and $C=\frac{141583}{82031250}=0.00172596$, which deviate only $1\%$ from the exact results, obtained numerically from \eqref{Q1}-\eqref{Q3}.

\subsection*{Diffusive wire with a barrier and NS-reservoirs}

Another interesting application of the scaling equation is a system, coupled to NS reservoirs, formed by the concatenation of an anomalous diffusive wire with a barrier of arbitrary transparency. We shall focus attention on the comparison between the behavior of the NS conductance and the conductance of the same system with normal metal reservoirs in order to highlight the role of tunneling in the presence of Andreev's reflection. 

The pseudo-current $I_s(\phi)\equiv I(\phi,N_{\text{ch}}s)$ for a system formed by the concatenating a barrier of transparency $\Gamma$ and an anomalous diffusive wire satisfies equation \eqref{sol_implicit} with initial condition given by
\begin{equation}
I_{0}(\phi)=\frac{N_{\text{ch}}\Gamma\sin\phi}{1-\Gamma\sin^2(\phi/2)}.
\end{equation}
We calculate the average NS conductance from the relation $\langle g_{\text{NS}}\rangle_{s} = F^{\prime}_{s}(\pi/2)$, where  $F_{s}(\phi)\equiv I_s(\phi)/\sin\phi$. Next, we determine the coefficients $x$ and $y$ of the series expansion of $F_s(\phi)$, in the form of $F_s(\phi)=x + y\,\delta\phi+\ldots$ for $\delta\phi\equiv\phi-\pi/2$, where $x=F_s(\pi/2)$ and $y=F^{\prime}_s(\pi/2)$. We find
\begin{equation}
g_{\text{NS}} = y = \frac{N_{\text{ch}}\,\Gamma}{\Gamma s + (Q/\Gamma)^{-1}},
\end{equation}
where $Q$, the effective tunneling coefficient,  is given by
\begin{equation}\label{eq_tunelamento}
\left(Q/\Gamma\right)^{-1}=\frac{\Gamma s\,\cos\beta}{\theta}\left[\,\left(\,1+\sin\beta\,\right)\frac{\theta}{\Gamma s\,\cos\beta}-1\,\right]^{-1}\\ + \frac{\Gamma s}{\cos(\beta-\theta)}\,\frac{\Theta(1-\lambda)}{\lambda},
\end{equation}
where the variables $\beta$ and $\theta\equiv s\,x$ are determined by the following system of transcendental equations
\begin{align}\label{12}
&\sin(\beta-\theta)=\theta\,\frac{\Theta(1-\lambda)}{\lambda}& &\text{and}&
&\theta\,\left[\,1-\frac{\Gamma}{2}\left(1-\sin\beta\right)\,\right]=\Gamma s\,\cos\beta.&
\end{align}

\begin{figure}[h!]\centering
\subfigure[ref1][\, $\lambda > 1.0$]{\includegraphics[width=0.35\textwidth]{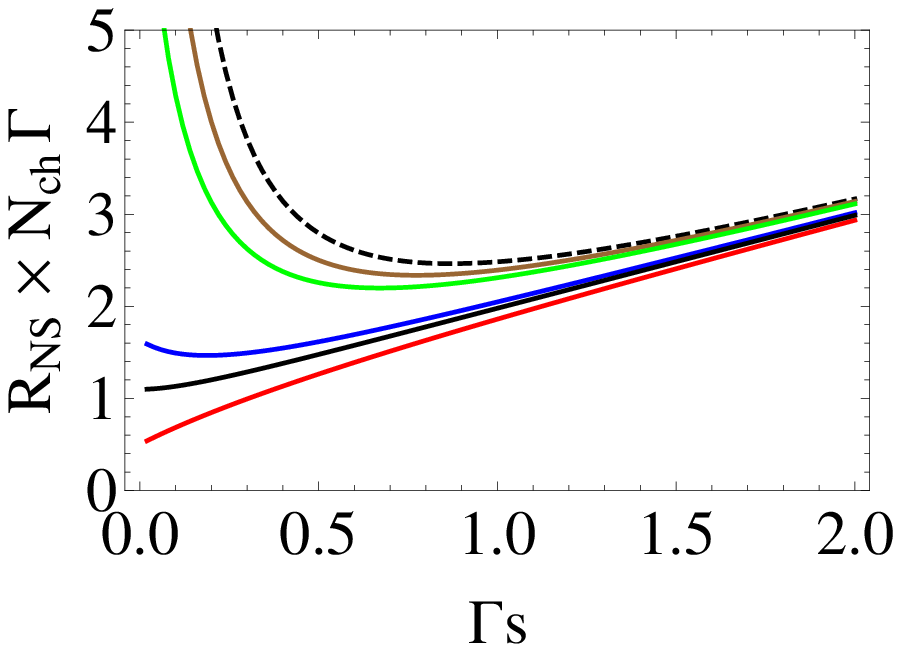}\label{2a}}\qquad
\subfigure[ref2][\, $\lambda = 1.0$]{\includegraphics[width=0.35\textwidth]{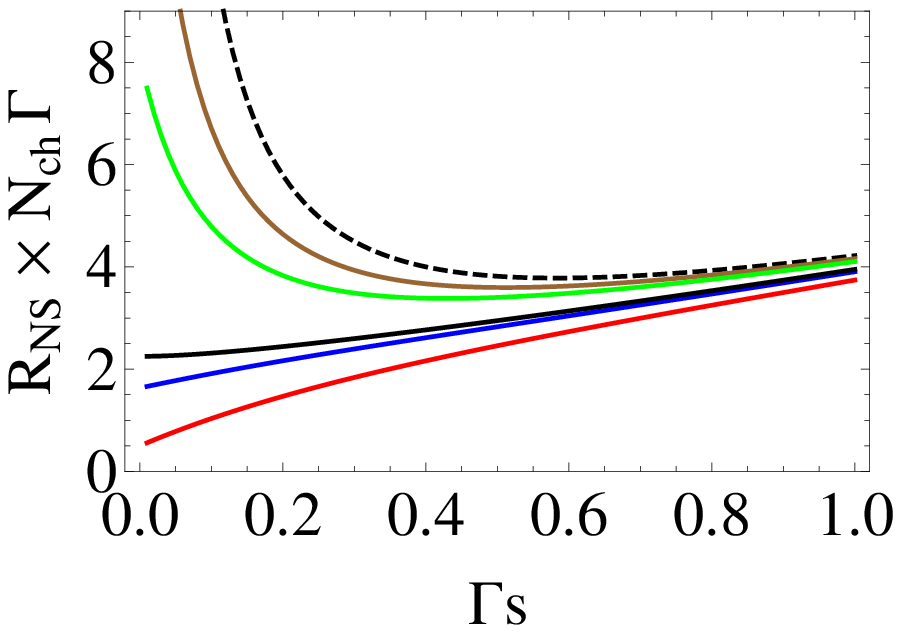}\label{2b}}\\
\subfigure[ref2][\, $\lambda = 0.5$]{\includegraphics[width=0.35\textwidth]{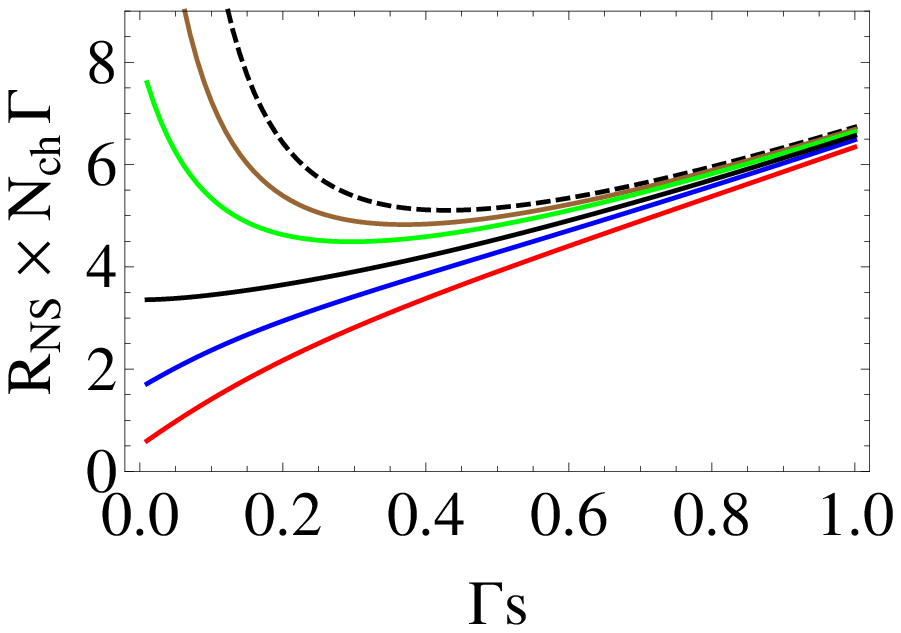}\label{2c}}\qquad
\subfigure[ref2][\, $\lambda = 0.1$]{\includegraphics[width=0.35\textwidth]{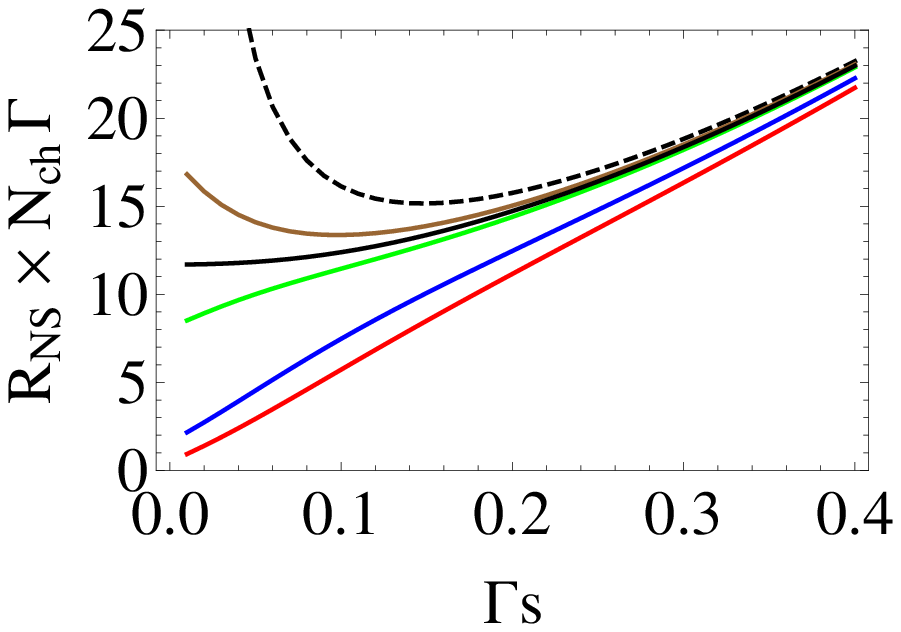}\label{2d}}
\caption{Resistance $R_{NS}\times N_{\text{ch}}\Gamma$ as a function of $\Gamma s$ for different values of transparency $\Gamma$ and suppression parameter $\lambda$. The dashed curve corresponds to $\Gamma\ll 1$. The brown, green, blue and red curves corresponds to $\Gamma=0.1,\, 0.2,\, 0.6\,\text{and}\,1.0$, respectively. The black curves correspond to $\Gamma=\Gamma_{\text{threshold}}$, which represent the separatrix curve of the change monotonicity, according to the equation \eqref{threshold} (color online). }
\label{fig:Fano_G_NS}
\end{figure}

Note that for $\lambda>1$, the above system is simplified since $\theta=\beta$. The dimensionless resistance, $R_{\text{NS}}\equiv(g_{\text{NS}})^{-1}$, was calculated numerically as a function of $\Gamma s$ from the above equations. In figure \ref{fig:Fano_G_NS}, we show $R_{\text{NS}}\times N_{\text{ch}}\,\Gamma$ as a function of $\Gamma s$. Remarkably, a minimum in the NS resistance, which is a distinctive signature of the reflectionless tunneling effect in usual NS systems, where $\lambda>1$ (see  \cite{Beenakker1997} for a review), is also present for anomalous NS systems, where $0<\lambda \leqslant 1$.  

We give below analytical expressions for $R_{\text{NS}}$ in the regimes $\Gamma s\gg 1$ or $\Gamma s\ll 1$ for arbitrary $\lambda$,
For $\Gamma s\gg 1$, we obtain
\begin{equation}
R_{\text{NS}}\times N_{\text{ch}}\Gamma\approx \begin{cases}
\Gamma s +1,\;\text{for}\;\lambda>1;\\
a(\lambda)\,\left[\,\Gamma s + q(\lambda)\,\right],\;\text{for}\; 0<\lambda\leqslant 1,
\end{cases}
\end{equation}
where $q(\lambda)=b(\lambda)/a(\lambda)$ and
\begin{align}
&a(\lambda)=1+\frac{1}{\lambda\sin\theta_{0}},&  
&b(\lambda)=1+\frac{\lambda(\lambda^{2}-\theta^{2}_{0}-\lambda)(1-\sin\theta_0)}{(\lambda^{2}-\theta_{0}^{2})(1+\lambda\sin\theta_{0})},&
\end{align}
where the parameter $\theta_{0}$ is determined by $\theta_{0}=\lambda\cos\theta_{0}$.\\

For $\Gamma s\ll 1$, we obtain
\begin{subequations}
\begin{equation}
R_{\text{NS}}\times N_{\text{ch}}\Gamma = \frac{(2-\Gamma)^2}{2\Gamma}+\frac{2\,H_{\lambda}(\Gamma)}{\Gamma^2}\left(\Gamma s\right) + \mathcal{O}\left[(\Gamma s)^2\right],
\end{equation}
where
\begin{equation} \label{H}
H_{\lambda}(\Gamma)\equiv\Gamma^2 + 2\Gamma -2 + (\Gamma^2+\Gamma)\frac{\Theta(1-\lambda)}{\lambda}.
\end{equation}
\end{subequations}
Note that for $s\ll 1$, the conductance of the NS-system is given by $g_{\text{NS}}\sim \Gamma$ if $\Gamma s\gg 1$ and $ g_{\text{NS}}\sim\Gamma^{2}$ if $\Gamma s\ll 1$. The change of behavior from $\Gamma^2$ to $\Gamma$ as one increases the size of the system signals the change of the effective tunneling process from two-particle (electron-hole) to single-particle (electron), indicating a reflectionless tunneling through the barriers of the Andreev reflected hole. For the usual NS system ($\lambda > 1$) this transition is explained by the disorder induced opening of ideal transmission channels \cite{Nazarov1994}. Since this mechanism is not justified for the anomalous metal, where the Fabry-Perot modes are suppressed ($0<\lambda\leqslant 1$), the observation of reflectionless tunneling in this case is indeed an interesting surprise.

The role of the FP suppression parameter, $\lambda$, can be best understood by extracting the  condition for the appearance of a \textit{resistance minimum} in the system. From \eqref{H} we find that we must have $H_{\lambda}(\Gamma_{\text{threshold}})<0$, which is equivalent to $\Gamma<\Gamma_{\text{threshold}}$, where
\begin{equation}\label{threshold}
\Gamma_{\text{threshold}}=\begin{cases}
\sqrt{3}-1, &\text{for}\quad\lambda>1;\\
\dfrac{-(1+2\lambda) +\sqrt{1+12\lambda+12\lambda^2}}{2 (1+\lambda)}, &\text{for}\quad 0<\lambda\leqslant 1,
\end{cases}
\end{equation}
which implies that $0<\Gamma_{\text{threshold}}\leqslant 1/2$ and that it increases monotonically in the interval $0<\lambda\leqslant 1$. Therefore, for a fixed value of the barrier transparency $0< \Gamma < 1/2$ we can completely suppress the reflectionless tunneling effect in the anomalous metal by reducing the value of $\lambda$, i.e. by continuously removing highly transmitting channels. Likewise, we can produce the effect by increasing the value of $\lambda$, provided $0< \Gamma < 1/2$.  Such a control mechanism is a unique feature of the anomalous metal NS junction, which may have interesting applications.


\section{Summary and Conclusions}

We presented a detailed derivation of a two-parameter scaling equation for the ballistic-anomalous-diffusive crossover under general conditions, which includes arbitrary shaped disordered conductors and quantum dot networks. The stationary solution of the two-parameter scaling equation generalizes Dorokhov's universal distribution of transmission eigenvalues for the dominant contribution in the semiclassical regime. We provided a simple interpretation of the stationary solution in terms of  the phase diagram and thermodynamics of a spin-glass system.  As an application, we considered a system formed by a diffusive wire coupled via a barrier to normal-superconductor (NS) reservoirs. We observed anomalous reflectionless tunneling, when all perfectly transmitting channels are suppressed, which cannot be accounted for by the usual mechanism of disorder-induced opening of perfectly transmitting channels. We showed that in this case the anomalous metal system offers a unique control mechanism for the emergence or suppression of reflectionless tunneling for fixed barrier transparencies in the interval $0 < \Gamma < 1/2$. 

\ack This work was partially supported by CNPq and CAPES (Brazilian Agencies). We are grateful to Ernesto P. Raposo for stimulating discussions and suggestions.


\appendix
\section{Proofs of Theorems and Corollaries}\label{Appen_A}

\subsection{Theorem 1}\label{Appen_A1}

We start by remarking that \eqref{kirchhoff} for $\phi=\pi-0^{+}$ can be rewritten as
\begin{equation}\label{equality}
I(\pi-0^{+})\equiv I_{L}(\xi)=\tilde{I}_{S}(\chi)=I_{R}(\xi^{\p}),
\end{equation}
where $\tilde{I}_{S}(\chi)$ is the pseudo-current of the sample, which according to \eqref{spectral_current} can be related to the transmission eigenvalue density, $\rho_{S}(\tau)$, through
\begin{subequations}\label{J_S}
\begin{equation}
\tilde{I}_{S}(\chi) \equiv\int_{0}^{1}\ud\tau\,\rho_{S}(\tau)\,\frac{2\,\tau\,\chi}{1+(1-\tau)\,\chi^2}= 2\,g_{S}\,\chi\,+\,2\,g_{S}\,J(\chi)\,\equiv I_{S}(\theta-\theta^{\p}),
\end{equation}
where $\chi=\tan(\theta/2-\theta^{\p}/2)$. We represent the function $J(\chi)$ as a power series in $\chi$ (for $\vert\chi\vert<1$) as
\begin{equation}
J(\chi)=\sum_{n=1}^{\infty}(-1)^{n}\,a_{n}\,\chi^{2n+1}\quad\text{and}\quad a_{n}=\frac{\,\int_{0}^{1}\ud\tau\,\rho_{S}(\tau)\,\tau\,(1-\tau)^{n}\,}{\int_{0}^{1}\ud\tau\,\rho_{S}(\tau)\,\tau},
\end{equation}
\end{subequations}
where $0<a_{n}<1$. Furthermore $I_{L}(\xi)$ and $I_{R}(\xi^{\p})$ are parametrized as in equations \eqref{I_LR}. Define the auxiliary variable $\chi=\left(\xi-\xi^{\p}\right)/\left(1+\xi\,\xi^{\p}\right)$, where $\xi=\tan(\theta/2)$ and $\xi^{\p}=\tan(\theta^{\p}/2)$. Note that $g_{S}=\partial_{\varphi}I_{S}(\varphi)\bigl\vert_{\varphi=0}=(1/2)\,\partial_{\chi}\tilde{I}_{S}(\chi)\bigl\vert_{\chi=0}=\int_{0}^{1}\ud\tau\,\rho_{S}(\tau)\,\tau$ is the conductance of the sample. Performing the change $J(\chi)\to\epsilon\,J(\chi)$ in \eqref{J_S}, we divide the proof of Theorem \ref{theor_1} in two parts: \textbf{(i)} we take $\epsilon=0$ and prove that this is sufficient to establish \eqref{zeta_LR}, provided $\xi=\xi_{0}$ and $\xi^{\p}=\xi^{\p}_{0}$ satisfy a polynomial equation obtained from \eqref{kirchhoff}; \;\;\textbf{\text{(ii)}} we express $\xi$ and $\xi^{\p}$ as a series in powers of $\epsilon$ as $\xi=\xi_{0} + \sum_{m\geqslant 1}\epsilon^{m}\,\xi_{m}$ and $\xi^{\p}=\xi_{0}^{\p} + \sum_{m\geqslant 1}\epsilon^{m}\,\xi_{m}^{\p}$ by \eqref{kirchhoff} and determine the conditions to establish \eqref{zeta_LR} for the coefficients $\lbrace a_{n}\rbrace$. In the final step, we take $\epsilon=1$.\\

\textbf{(i)} Solving equation \eqref{equality} with the aid of equation \eqref{I_LR} and \eqref{J_S} for $\epsilon=0$, we obtain a cubic equation in the variable $\xi^{2}$ 
\begin{align}\label{P}
&P(\xi_{0})\equiv\sum_{k=0}^{3} p_{k}\,\xi^{2k}_{0}= 0,&
\end{align}
where the coefficients $\lbrace p_{k}\rbrace$ are given by
\begin{subequations}
\begin{align}
p_{0}=&g_{S}^{2}\,g_{L}\,g_{R}\,(1-T_{L})^{2}\left(\frac{1}{N_L}+\frac{1}{g_S}+\frac{1}{g_R}-\frac{1}{g_L}\right),&\label{p0}\\
p_{1}=&g_{L}^3 (1-T_{R}) + g_{L}^2 (1-T_{L}) (g_{R}+2 g_{S} T_{R})-3\, g_{R}\,g_{S}^2 (1-T_{L})^2 &\nonumber\\
&\qquad\quad -g_{L}\,g_{S} (1-T_{L})\,\left[g_{R} (1+T_{L}) + g_{S} (3 - T_{R} -T_{L} + T_{L}\, T_{R})\right],&\label{p1}\\
p_{2}=&g_{L}^3+g_{L}^2 (g_{R} + 2\,g_{S}\,T_{R})-3 g_{R}\,g_{S}^2 (1-T_{L})-g_{L}\,g_{S}\,g_{R}\,\left(1-2 T_{L}\right)\nonumber\\
&\qquad\quad + g_{L} g_{S}^{2} \left[\,3-2 \left(T_{L}+T_{R}-T_{L} T_{R}\right)\,\right],&\label{p2}\\
p_{3}=&-g_{S}^{2}\,g_{R}\,g_{L}\left(\frac{1}{N_{R}}+\frac{1}{g_{S}}+\frac{1}{g_{L}}-\frac{1}{g_{R}}\right).\label{p3}&
\end{align}
\end{subequations}
The physical root of equation \eqref{P}, $\xi_{0}=\xi_{\text{sol}}^{(0)}$, is the one which gives $\nu(0)\geqslant 0$. The boundary of the support of the FP modes, $\nu(0)=0^{+}$, in the $T_{L}-T_{R}$ plane is given by $\text{Re}\left[I_{L}(\xi_{0})\right]=\text{Re}\left[I_{R}(\xi^{\p}_{0})\right]=\text{Re}\left[I_{S}(\chi_{0})\right]=0^{+}$ (where $\chi_{0}=(\xi_{0}-\xi^{\p}_{0})/(1+\xi_{0}\,\xi^{\p}_{0})$), which in turn corresponds to the choice $\xi_{0}=0^{+}=\xi^{\p}_{0}$, or $p_{0}=0^{+}$, in equation \eqref{P}. From \eqref{I_LR} we observe that the exchange $L\leftrightarrow R$ is equivalent to setting $\xi_{0}\leftrightarrow 1/\xi^{\p}_{0}$ in equation \eqref{P}, which implies $\sum_{k=0}^{3}p_{3-k}\,\xi^{\p\,2k}_{0}=0$. Thus, if we set $\xi^{\p}_{0}= 0^{+}$, we find $p_{3}=0^{+}$.
Using $p_{0}=0^{+}=p_{3}$ in \eqref{p0} and \eqref{p3}, provides the boundary condition $\zeta=\zeta_{c}-0^{+}$ in agreement with \eqref{zeta_LR}.\\

\textbf{(ii)} We must show that for any function $J(\chi)$, or equivalently for any set $\left\lbrace a_{k}\right\rbrace$,  equation \eqref{zeta_LR} is still valid. Let us represent the auxiliary variables $\xi$ and $\xi^{\p}$ as
\begin{align}\label{EqA5}
&\xi=\xi_{0}+\sum_{m\geqslant 1}\epsilon^{m}\,\xi_{m}& &\text{and}& &
\xi^{\p}=\xi^{\p}_{0}+\sum_{m\geqslant 1}\epsilon^{m}\,\xi^{\p}_{m},&
\end{align}
and  express $I_{L}(\xi)$ and $I_{R}(\xi^{\p})$, using \eqref{I_LR} and the auxiliary variable $\chi=(\xi-\xi^{\p})/(1+\xi\,\xi^{\p})$, as 
\begin{subequations} 
\begin{align}
&I_{L}(\xi)=I_{L}(\xi_{0})+2\,g_{L}\sum_{m\geqslant 1}\epsilon^{m}\,\left[\,\xi_{m}\,A_{L}(\xi_{0})+B^{(m)}_{L}(\vec{\xi}_{[m]})\,\right],\quad\text{with}\quad B^{(1)}_{L}\equiv 0&\label{IL_e}\\
&I_{R}(\xi^{\p})=I_{R}(\xi_{0}^{\p})+2\,g_{R}\sum_{m\geqslant 1}\epsilon^{m}\,\left[\,\xi_{m}^{\p}\,A_{R}(\xi_{0}^{\p})+B^{(m)}_{R}(\vec{\xi}^{\,\p}_{[m]})\,\right],\quad\text{with}\quad B^{(1)}_{R}\equiv 0&\label{IR_e},
\end{align}
\end{subequations}
where we define the two $m$-uple $\vec{\xi}_{[m]}\equiv(\xi_{0},\ldots,\xi_{m-1})\,$ and $\,\vec{\xi}^{\,\p}_{[m]}\equiv(\xi_{0}^{\p},\ldots,\xi_{m-1}^{\p})$. Moreover
\begin{align}
&A_{L}(\xi_{0})=\frac{1-T_{L}-\xi_{0}^{2}}{\left(1-T_{L}+\xi_{0}^{2}\right)^{2}}& &\text{and}& &A_{R}(\xi_{0}^{\p})=\frac{1-(1-T_{R})\,\xi^{\p\, 2}_{0}}{\left[\,1+(1-T_{R})\,\xi^{\p\,2}_{0}\,\right]^{2}},&
\end{align}
and also
\begin{equation}\label{BL}
B^{(m)}_{L}(\vec{\xi}_{[m]})=\sum^{m}_{\substack{
   a_{1},\,a_{2},\,\ldots\, ,\,a_{m-1} = 0 \\
   \left(\sum_{j=1}^{m-1}j\,a_{j}=m\right)
  }}\xi_{1}^{a_{1}}\,\xi_{2}^{a_{2}}\,\cdots\,\xi_{m-1}^{a_{m-1}}\,\,b_{L}^{\mathcal{P}}(\xi_{0}),\quad\text{for}\quad m\geqslant 2,
\end{equation}
where we introduced the following notation  $\mathcal{P}\equiv\left[a_{1},a_{2},\ldots, a_{m-1}\right]$ for set of $2(m-2) + \delta_{m,2}$ functions $\left\lbrace b_{L}^{\mathcal{P}}(\xi_{0})\right\rbrace$. For $\vert u\vert\ll 1$, the set represented by $b_{L}^{\mathcal{P}}(u\,\xi_{0})$ are homogeneous functions. One can verify that
\begin{equation}\label{BLu}
B^{(m)}_{L}(u\,\vec{\xi}_{[m]})=-\frac{u^{3}}{\left(1-T_{L}\right)^2}\sum^{m-1}_{\substack{
   i, j, k\,=\,0 \\
   \left(i + j + k\,=\,m\right)
  }} \xi_{i}\,\xi_{j}\,\xi_{k}\,\,+\,\, \mathcal{O}(u^5),
\end{equation}
Similarly to \eqref{BL}, we define $B^{(m)}_{R}(\vec{\xi^{\p}}_{[m]})$ by $\xi_{j}\to\xi^{\p}_{j}$ and $\,b_{L}^{\mathcal{P}}(\xi_{0})\to b_{R}^{\mathcal{P}}(\xi^{\p}_{0})\,$. Similarly to \eqref{BLu}, we established
\begin{equation}\label{BRu}
B^{(m)}_{R}(u\,\vec{\xi^{\p}}_{[m]})=-\,u^{3}\left(1-T_{R}\right)\sum^{m-1}_{\substack{
   i, j, k\,=\,0 \\
   \left(i + j + k\,=\,m\right)
  }} \xi_{i}^{\p}\,\xi_{j}^{\p}\,\xi_{k}^{\p}\,\,+\,\, \mathcal{O}(u^5).
\end{equation}

In order to expand $\tilde{I}_{S}(\chi)$ in a series of powers of $\epsilon$, expand $\chi^{2n+1}$ as follows
\begin{multline}\label{chi}
\frac{\chi^{2n+1}}{\chi_{0}^{2n+1}}=1+(2n+1)\sum_{m\geqslant 1}\epsilon^{m}\,\left[\,\xi_{m}\,X(\xi_{0},\xi_{0}^{\p})+\xi_{m}^{\p}\,Y(\xi_{0},\xi_{0}^{\p})+Z_{n}^{(m)}(\vec{\xi}_{[m]},\,\vec{\xi^{\p}}_{[m]};n)\,\right],
\end{multline}
where $Z^{(1)}_{n}\equiv 0$
and $\chi_{0}\equiv(\xi_{0}-\xi_{0}^{\p})/(1+\xi_{0}\,\xi_{0}^{\p})$ and also 
\begin{subequations}
\begin{align}
&X(\xi_{0},\xi_{0}^{\p})=\frac{1+\xi^{\p\, 2}_{0}}{\left(\xi_{0}-\xi^{\p}_{0}\right)\left(1+\xi_{0}\,\xi_{0}^{\p}\right)},& &Y(\xi_{0},\xi_{0}^{\p})=-\frac{1+\xi^{2}_{0}}{\left(\xi_{0}-\xi_{0}^{\p}\right)\left(1+\xi_{0}\,\xi_{0}^{\p}\right),}&
\end{align}
\end{subequations}
where
\begin{equation}\label{Z_n^m}
Z^{(m)}_{n}(\vec{\xi}_{[m]},\,\vec{\xi^{\p}}_{[m]};n)=\sum_{a=1}^{m}n^{m-a}\,z_{a}^{(m)}(\vec{\xi}_{[m]},\,\vec{\xi^{\p}}_{[m]}),
\end{equation}
and the set functions $z^{(m)}_{m}, z^{(m)}_{m-1}, \ldots, z^{(m)}_{1}$ for $m\geqslant 2$ exhibit a leading behavior for $\xi_{a}\to u\,\xi_{a}$ and $\xi^{\p}_{a}\to u\,\xi^{\p}_{a}$ with $\vert u\vert\ll 1$  as
\begin{subequations}
\begin{align}
&z^{(m)}_{m}(u\,\vec{\xi}_{[m]},\,u\,\vec{\xi^{\p}}_{[m]})=u^{2}\,G_{m}^{(m)}(\vec{\xi}_{[m]},\,\vec{\xi^{\p}}_{[m]}) + \mathcal{O}(u^4),\label{A14a}\\
& z^{(m)}_{a}(u\,\vec{\xi}_{[m]},\,u\,\vec{\xi^{\p}}_{[m]})=G^{(m)}_{a}(\vec{\xi}_{[m]},\,\vec{\xi^{\p}}_{[m]}) + \mathcal{O}(u^{2}),\quad\text{for}\quad a=0,\,\ldots,\,m-1,\label{A14b}
\end{align}
\end{subequations}
where the function $G_{m}^{(m)}$ is defined as
\begin{equation}
G_{m}^{(m)}(\vec{\xi}_{[m]},\,\vec{\xi^{\p}}_{[m]})\equiv\sum_{\substack{
   i,\,j,\,k\,=\,0\\
   \left(i+j+k=m\right)
  }}^{m-1} (-1)^{i j k + \Delta^{(m)}_{i,j,k}} \,\,\frac{\,\xi^{\p}_{i}\,\xi^{\p}_{j}\,\xi_{k} - \xi_{i}\,\xi_{j}\,\xi^{\p}_{k}\,}{\xi_{0}-\xi^{\p}_{0}},\label{Hm}
\end{equation}
where $\Delta^{(m)}_{i,j,k}=\delta_{i,1}\,\delta_{j,1}+ \delta_{i,1}\,\delta_{k,1} + \delta_{j,1}\,\delta_{k,1}$ (for $m$ odd) and $\Delta^{(m)}_{i,j,k}\equiv 0$ (for $m$ even). The set of functions $\big\lbrace G_{a}^{(m)}\big\rbrace_{0\leqslant a\leqslant m-1}$ can be defined for the first values of $m=2, 3, 4, \ldots$ as 
\begin{subequations}
\begin{align}
G^{(2)}_{1}&\equiv\Gamma^{2}_{1,2}\, ,\\
G^{(3)}_{1}&\equiv\frac{2}{3}\,\Gamma^{3}_{1,3}\, ,&
G^{(3)}_{2}&\equiv\frac{1}{3}\,\Gamma^{3}_{2,3}\, ,\\
G^{(4)}_{1}&\equiv\frac{1}{3}\,\Gamma^{4}_{1,4}\, ,& 
G^{(4)}_{2}&\equiv -\frac{1}{2}\Gamma^{4}_{1,4} + 2\,\Gamma^{2}_{1,2}\,\Gamma^{2}_{2,1}\, ,& 
G^{(4)}_{3}&\equiv\Gamma^{4}_{1,4}-\frac{1}{3}\,\Gamma^{4}_{3,3}\, ,\quad\text{etc},
\end{align}
\end{subequations}
where the symbols $\Gamma^{s}_{r, n}$ are defined in terms of
\begin{align}
&\Gamma^{s}_{r,n}\equiv\sum_{k=0}^{n}(-1)^{r+k}\,\binom {n}{k}\left[\sum_{\scriptscriptstyle i_{1},\ldots, i_{n-k}=0}^{r}\,\sum_{\scriptscriptstyle j_{1},\ldots j_{k}=0}^{r}\,(-1)^{\pi+\sigma_{s}}\,\delta_{\mu,s}\,\frac{\prod_{\alpha=1}^{n-k}\xi_{i_{\alpha}}\prod_{\beta=1}^{k}\xi^{\prime}_{j_{\beta}}}{\left(\xi_{0}-\xi_{0}^{\p}\right)^{n}}\right],
\end{align}
Also, we define the symbols
$\pi\equiv\prod_{x=1}^{n-k}i_{x}\,\prod_{y=1}^{k} j_{y}\,$, $\mu\equiv\sum_{x=1}^{n-k}i_{x}+\sum_{y=1}^{k} j_{y}$ and $\sigma_{s}=\big(\sum_{a=1}^{n-k}\prod_{x=1\,(x\neq a)}^{n-k}\delta_{i_{x},1}\big)\big(\prod_{y=1}^{k}\delta_{i_{y},1}\big) + \big(\prod_{x=1}^{n-k}\delta_{i_{x},1}\big)\big(\sum_{b=1}^{k}\prod_{y=1\,(y\neq b)}^{k}\delta_{i_{y},1}\big)$ (for $s$ even) or $\sigma_{s}\equiv 0$ (for $s$ odd). Note that we impose the condition $2\leqslant s\leqslant r\,n$. For example, it is possible to write the condition represented by the equality ($s=r\,n$) as $\Gamma^{r\,n}_{r,\, n}=\left(\xi_{r}-\xi^{\p}_{r}\right)^{n}/\left(\xi_{0}-\xi^{\p}_{0}\right)^{n}$. Substituting equations \eqref{A14a}-\eqref{A14b} in  \eqref{Z_n^m}, we obtain the leading terms $Z^{(m)}_{n\neq 0}(u\,\vec{\xi}_{[m]},\,u\,\vec{\xi^{\p}}_{[m]}; n) = \sum_{a=1}^{m-1} n^{m-a}\,G^{(m)}_{a}(\vec{\xi}_{[m]},\,\vec{\xi^{\p}}_{[m]}) + \mathcal{O}(u^2)$ and $Z^{(m)}_{n=0}(u\,\vec{\xi}_{[m]},\,u\,\vec{\xi^{\p}}_{[m]}; 0)= u^{2}\,G_{m}^{(m)}(\vec{\xi}_{[m]},\,\vec{\xi^{\p}}_{[m]})+\mathcal{O}(u^4)$.

From \eqref{J_S} and subsequent results, we can write $\tilde{I}_{S}(\chi)$ as
\begin{equation}\label{IS_e}
\tilde{I}_{S}(\chi)=2\,g_{S}\,\chi_{0}
+2 g_{S}\,\sum_{m\geqslant 1}\epsilon^{m}\left(\,\xi_{m}\,\check{X}+\xi^{\p}_{m}\,\check{Y} + W^{(m)}\,\right),
\end{equation}
where $\check{X}(\xi_{0},\xi_{0}^{\p})\equiv\chi_{0}\,X(\xi_{0},\xi_{0}^{\p})$, $\check{Y}(\xi_{0},\xi_{0}^{\p})\equiv\chi_{0}\,Y(\xi_{0},\xi_{0}^{\p})$ and $W^{(m)}=W^{(m)}(\vec{\xi}_{[m]},\,\vec{\xi^{\p}}_{[m]})$, with
\begin{subequations}
\begin{align}
&W^{(1)}\equiv J(\chi_{0}),\label{Wm1}\\
&W^{(m\geqslant 2)}\equiv\,\left(\xi_{m-1}\,\check{X}+\xi^{\p}_{m-1}\,\check{Y}\right)\,\partial_{\chi_{0}}J(\chi_{0})+\sum_{n=1}^{\infty}(-1)^{n}\,a_{n}\,(2n+1)\,\chi_{0}^{2n+1}\,Z_{n}^{(m-1)}+\chi_{0}\,Z^{(m)}_{n=0}.\label{Wm2}\qquad\qquad
\end{align}
\end{subequations}

Substituting \eqref{Z_n^m}, \eqref{A14a} and \eqref{A14b} in \eqref{Wm1} and \eqref{Wm2}, we expressed the leading behavior of 
$W^{(m)}(u\,\vec{\xi}_{[m]},u\,\vec{\xi}_{[m]}^{\,\p})$ for $u=0^{+}$ as
\begin{multline}
W^{(m)}(u\,\vec{\xi}_{[m]},u\,\vec{\xi}_{[m]}^{\,\p})=-u^{3}\,\Big[3\,a_{1}\left(\xi_{0}-\xi_{0}^{\p}\right)^{2}\left(\xi_{m-1}-\xi_{m-1}^{\p}\right)\,+\,\Theta(m-2)\,\left(\xi_{0}-\xi^{\p}_{0}\right)\\
\times\Big(\,G^{(m)}_{m}(\vec{\xi}_{[m]},\vec{\xi}_{[m]}^{\,\p})-3\,a_{1}\sum_{a=1}^{m-2}G^{(m-1)}_{a}(\vec{\xi}_{[m-1]},\vec{\xi}_{[m-1]}^{\,\p})\,\Big)\Big]+\mathcal{O}(u^5).
\end{multline}

Inserting \eqref{IL_e}, \eqref{IR_e} and \eqref{IS_e} into \eqref{equality}, we get the following set of equations for the set $\lbrace\xi_{m},\xi^{\p}_{m}\rbrace$ 
\begin{subequations}\label{A41ab}
\begin{gather}
I_{L}(\xi_{0})=2\,g_{S}\,\chi_{0}=I_{R}(\xi_{0}^{\p}),\quad\text{for zero order in}\;\epsilon\label{xi_0}\\
g_{L}\,\left(\xi_{m}\,A_{L}+B_{L}^{(m)}\right)= g_{S}\,\left(\xi_{m}\,\check{X}+\xi^{\p}_{m}\,\check{Y} + W^{(m)}\,\right)=g_{R}\left(\xi^{\p}_{m}\,A_{R} + B^{(m)}_{R}\right),\,\,\text{for}\,\,m\geqslant 1.\label{xi_n}
\end{gather}
\end{subequations}
The solution of \eqref{xi_0} is then given by $P(\xi_{0})=0$ in agreement with \eqref{P}. Solving \eqref{xi_n} for $(\xi_{m},\xi^{\p}_{m})$ with $m\geqslant 1$, we find
\begin{align}\label{xi_sistem}
\xi_{m}=\frac{\mathcal{Y}^{(m)}}{\mathcal{W}^{(m)}}\qquad\text{and}\qquad\xi_{m}^{\p}=\frac{\mathcal{X}^{(m)}}{\mathcal{W}^{(m)}}\, ,
\end{align}
where
\begin{subequations}
\begin{align}
&\mathcal{Y}^{(m)} \equiv g_{R}\,A_{R}\,\big(g_{L}\,B^{(m)}_{L} -  g_{S}\,W^{(m)}\big) + g_{S}\,\big(\,g_{R}\,B_{R}^{(m)} - g_{L}\,B_{L}^{(m)}\,\big) \check{Y},\label{Ym}\\
&\mathcal{X}^{(m)} \equiv g_{L}\,A_{L}\big(\,g_{R}\,B^{(m)}_{R}-g_{S}\,W^{(m)}\,\big) + g_{S}\,\big(\,g_{L}\,B_{L}^{(m)} - g_{R}\,B_{R}^{(m)}\,\big)\,\check{X},\label{Xm}\\ 
&\mathcal{W}^{(m)} \equiv g_{S}\,\big(\,g_{L}\,A_{L}\,\check{Y} + g_{R}\,A_{R}\,\check{X}\,\big)-g_{L}\,g_{R}\,A_{L}\,A_{R}.\label{Wm}
\end{align}
\end{subequations}

For $\zeta=\zeta_{c} - 0^{+}$ and $g_{L}<g_{R}$ (or $p_{0}=0^{+}$ according to \eqref{p0}), we get $\xi_{0}=0^{+}=\xi^{\p}_{0}$. Therefore, for  $\zeta=\zeta_{c}\,(1 - t)$, $\xi_{0} = u\,y_{0}$ and $\xi_{0}^{\p} = u\,x_{0}$ with $u=t^{\beta}=0^{+}$ and also for $y_{0}$ and $x_{0}$ finite, we obtain $\mathcal{Y}^{(m)}\propto u^{3}$, $\mathcal{X}^{(m)}\propto u^{3}$ (thus $B^{(m)}_{L,R}\propto u^{3}$ and $W^{(m)}\propto u^{3}$) and $\mathcal{W}^{(m)}\propto t = u^{1/\beta}$, where we use $\xi_{a} = u\,y_{a}$ and $\xi^{\p}_{a} = u\,x_{a}$ for $1\leqslant a\leqslant m-1$ and a finite set $\lbrace y_{a}, x_{a}\rbrace$. This assumption is consistent with \eqref{xi_sistem} with $\xi_{m},\xi_{m}^{\,\p} \propto u^{3-1/\beta}=u$ for $\beta=1/2$. By induction, we find $\xi_{m} = u\,y_{m}$ and $\xi^{\p}_{m} = u\,x_{m}$ with $m\geqslant 1$ for $u=0^{+}$ and $y_{m}$ and $x_{m}$ finite. Thus, we showed that $\xi_{m}=0^{+}=\xi^{\p}_{m}$, $\forall\,m$,  and we established from \eqref{EqA5} that $\xi=0^{+}$ and $\xi^{\p}=0^{+}$ for $\epsilon>0$.  

We are now in position to set $\epsilon=1$ and finnish the proof. We have therefore demonstrated that for the condition $\zeta=\zeta_{c}-0^{+}$, we have $\xi=0^{+}=\xi^{\p}$ and thus $\nu(0)=0^{+}$, as required by Theorem \ref{theor_1}.

\subsection{Corollary 1}\label{Appen_A2}

The parameter $\zeta$ shown in \eqref{zeta_LR} can be written as
\begin{equation}\label{zetaL}
\zeta\equiv\zeta_{R}\,\Theta(g_{L}-g_{R})\, +\, \zeta_{L}\,\Theta(g_{R}-g_{L})\quad\text{with}\quad\zeta_{\s} = N_{\s}\,\left(\frac{1}{g_{\s}} -\frac{1}{g_{\bar{\s}}}\right)
\end{equation}
for $\lbrace\s,\bar{\s}\rbrace = L, R$. Then, for $g_{R}>g_{L}$ we represent the boundary of the support of FP modes, i.e. the points where $\nu(0)=0^{+}$, as 
\begin{equation}\label{zeta_L}
\zeta_{L}=\left(1+\frac{1}{g_{S}/N_{L}}\right)(1-t),\quad\quad\text{therefore}\quad\quad g_{L}(t)=\frac{g_{R}\,g_{S}\left[\,1-(1-t)\,T_{L}\,\right]}{g_{S}+g_{R}\,(1-t)},
\end{equation}
for $t=0^{+}$. To obtain $\nu(0)$, we expand $\xi$ around $t=0^{+}$ in the form of \eqref{EqA5}, where
\begin{equation}
\xi^{2}_{0}(t)=\partial_{t}\,\xi_{0}^{2}(t)\Big\vert_{t=0}\,t + \mathcal{O}(t^{2}),\qquad\text{with}\qquad \partial_{t}\,\xi_{0}^{2}(t)\Big\vert_{t=0}=-\frac{\partial_{t}\,p_{0}(t)}{p_{1}(t)}\Biggl\vert_{t=0}\equiv y_{0}^2 > 0, 
\end{equation}
in which we used the relation $P(\xi_{0})=0^{+}$, defined in \eqref{P} for $t=0$, with $p_{k}=p_{k}(t)$ when $g_{L}=g_{L}(t)$. Thus, we obtain $\xi_{0}=y_{0}\,t^{\beta}$ with $\beta=1/2$ and $t=0^{+}$. Therefore, we have $\xi=t^{1/2}\,y$ and $\xi^{\p}=t^{1/2}\,x$, where $y\equiv y_{0} + \sum_{m\geqslant 1}\epsilon^{m}\,y_{m}$ and $x\equiv x_{0} + \sum_{m\geqslant 1}\epsilon^{m}\,x_{m}$.

We can obtain $\nu(0)$ using equation \eqref{nu}. We get
\begin{subequations}
\begin{equation}
\nu(0)=\frac{1}{\pi}\,\text{Re}\Big[I_{L}(\xi)\Big]=\mathcal{B}_{L}\,t^{1/2}\,\Theta(t),
\end{equation}
where the non-universal amplitude $\mathcal{B}_{L}$ is given by
\begin{equation}\label{bL}
\mathcal{B}_{L}= \mathcal{B}_{L}^{(0)} + \sum_{m\geqslant 1}\epsilon^{m}\,\mathcal{B}_{L}^{(m)}\, ,\quad\text{with}\quad\mathcal{B}_{L}^{(m)}=\frac{2}{\pi}\frac{g_{R}\,g_{S}}{\left(g_{R}+g_{S}\right)}\,\text{Re}\left[y_{m}\right],
\end{equation}
\end{subequations}
where the first term of the amplitude, $\mathcal{B}_{L}^{(0)}$, is
\begin{equation}
\mathcal{B}_{L}^{(0)}\equiv\frac{2}{\pi}\,g_{R}\,g_{S}\,\sqrt{\frac{g_{R}+g_{S}\,T_{L}}{g_{R}^{3}+g_{R}\, g_{S}\,(g_{R}+g_{S})\,(2+T_{L})+g_{S}^{3}\,(T_{L}+T_{R}-T_{L}T_{R})}}\,\, .
\end{equation}

The series shown in \eqref{bL} has guaranteed convergence for $\epsilon=1$ by virtue of the convergence of $y\vert_{\epsilon=1}=y_{0}+\sum_{m\geqslant 1}y_{m}$ by hypothesis.

The behavior of $\nu(0)$ at $\zeta_{R}=\zeta_{c} - 0^{+}$ can be obtained through the exchange $L\leftrightarrow R$. This is justified from the change of variables $\theta\to\phi-\theta^{\p}$ and $\theta^{\p}\to\phi-\theta$ in \eqref{kirchhoff}, since $I(\phi)\equiv I(\phi-\theta; T_{R})=I_{S}(\theta-\theta^{\p})=I(\theta^{\prime};T_{L})$, which ensures the symmetry $L\leftrightarrow R$. We may thus write $\nu(0) = \mathcal{B}\, t^{1/2}$ for $\zeta=\zeta_{c}-0^{+}$ and $\nu(0)=0$ for $\zeta>\zeta_{c}$ with $t\equiv (\zeta_{c}-\zeta)/\zeta_{c}$, where $\mathcal{B}=\mathcal{B}_{R}\,\Theta(g_{L}-g_{R})+\mathcal{B}_{L}\,\Theta(g_{R}-g_{L})$ (for $\epsilon=1$), which completes the proof of Corollary \ref{coro_1}. Note that if $g_{S}\to\infty$, we recover the result of a ballistic quantum dot with two barriers.

\subsection{Corollary 2}\label{Appen_A3}

From Theorem 1, the conservation of pseudo-current, shown in equation \eqref{kirchhoff}, can be written for $\phi= -\,2\,\ii\,x + \pi - 0^{+}$ and $\zeta=\zeta_{c}$ (see equation \eqref{zeta_LR}) as follows
\begin{equation}\label{eq_K}
K(\eta)\equiv K_{L}(\eta,\bar{\xi}) = K_{S}(\bar{\chi}) = K_{R}(\bar{\xi}^{\p}),
\end{equation}
where $K(\eta)$ is an analytic continuation of the pseudo-current $I(\phi)$ in the form $K(\eta)=-\ii\,I(\phi)$ with $\ii\,\eta = \cot(\phi/2) = \ii\,\tanh(x)+0^{+}$. Also the pseudo-currents of the $\s$-connectors can be rewritten as
\begin{subequations}\label{LRS}
\begin{align}
&K_{L}(\eta,\bar{\xi})\equiv\frac{2\,g_{L}\,(\eta+\bar{\xi})(1+\eta\,\bar{\xi})}{(1-\eta^2)(1-\bar{\xi}^2)-T_{L}(1+\eta\,\bar{\xi})^2} = -\ii\, I(\phi - \theta, T_{L}),\quad\text{with}\quad \ii\,\bar{\xi}=\tan(\theta/2),\label{KKL}\\
&K_{R}(\bar{\xi}^{\p})\equiv \frac{2\,g_{R}\,\bar{\xi}^{\p}}{1-(1-T_{R})\,\bar{\xi}^{\p\,2}} = - \ii\,I(\theta^{\p},T_{R}),\quad\text{with}\quad \ii\,\bar{\xi}^{\p}=\tan(\theta^{\p}/2).\label{KKR}
\end{align}
The pseudo-current, $K_{S}(\bar{\chi})$, of the sample is given by
\begin{equation}\label{KS}
K_{S}(\bar{\chi}) \equiv -\ii\,\tilde{I}_{S}(\ii\,\bar{\chi})  = 2\,g_{S}\left[\,\bar{\chi} + R(\bar{\chi})\,\right] \equiv -\ii\, I_{S}(\theta - \theta^{\p}),\;\;\text{with}\;\;\ii\,\bar{\chi} = \chi = \ii\,\frac{\bar{\xi}-\bar{\xi}^{\p}}{1-\bar{\xi}\,\bar{\xi}^{\p}},
\end{equation}
where $R(\bar{\chi}) \equiv -\ii\,J(\ii\,\bar{\chi})=\sum_{n=1}^{\infty} a_{n}\,\bar{\chi}^{2n+1}$, the latter equality follows from \eqref{J_S}.\\
\end{subequations}

The determination of $\nu(x)$ for $\zeta=\zeta_{c}$ (``critical isotherm'') proceeds from \eqref{nu}, since $\nu(x)=(-1/\pi)\,\text{Im}\left[K(\eta)\right]$ for $\eta=\tanh(x)$ and $\zeta=\zeta_{c}$. The realization of the condition $\zeta=\zeta_{c}$ for $g_{L}<g_{R}$ is, according to \eqref{zeta_LR}, obtained by inserting $g_{L}=g_{R}\,g_{S}\,(1-T_{L})/(g_{S}+g_{R})$ in \eqref{KKL}. The roots $\bar{\xi}$ and $\bar{\xi}^{\p}$ were obtained by following  a strategy similar to that adopted in \ref{Appen_A1}, after performing the change $R(\bar{\chi})\to\epsilon\,R(\bar{\chi})$ in \eqref{KS}. We divide the proof in two parts: \textbf{(i)} initially we choose $\epsilon=0$ and see that this is sufficient to establish the content of Corollary 2 from the determination of $\bar{\xi}=\gamma_{0}$ and $\bar{\xi}^{\p}=\gamma_{0}^{\p}$; \textbf{(ii)} we expand the roots as $\bar{\xi}=\gamma_{0} + \sum_{m\geqslant 1}\epsilon^{m}\,\gamma_{m}$ and $\bar{\xi}^{\p}=\gamma_{0}^{\p} + \sum_{m\geqslant 1}\epsilon^{m}\,\gamma_{m}^{\p}$ and find the condition to establish Corollary 2 by determination of the set $\lbrace\gamma_{m},\gamma_{m}^{\p}\rbrace$ for an arbitrary set $\lbrace a_{n}\rbrace$. At the end, we set $\epsilon=1$.\\

\textbf{(i)} The system of equations in \eqref{eq_K} becomes for $\epsilon=0$ the following: $K_{L}(\eta,\gamma_{0}) =2\,g_{S}\,\bar{\chi} = K_{R}(\gamma_{0}^{\p})$ (see \eqref{KKL}-\eqref{KS}), where  $\bar{\chi}=(\gamma_{0}-\gamma_{0}^{\p})/(1-\gamma_{0}\,\gamma_{0}^{\p})+\mathcal{O}(\epsilon)$. Therefore, we write $\xi_{0}^{\p}$ in terms of $\xi_{0}$ as
\begin{equation}\label{gamma0p}
\gamma_{0}^{\p}=\frac{2\,g_{S}\,\gamma_{0}-K_{L}(\eta,\gamma_{0})}{2\,g_{S}-\gamma_{0}\,K_{L}(\eta,\gamma_{0})},
\end{equation}
which means that we can establish a polynomial equation of 8-th degree for $\gamma_{0}$, denoted by $Q(\eta,\gamma_{0})=0$, where
\begin{equation}
\begin{split}\label{eq_xi0}
Q(\eta,\gamma_{0})\equiv&\left(g_R+g_S\right) \left[\left(1+\eta\,\gamma_0\right)^2\,T_L-\left(1-\eta ^2\right) \left(1-\gamma_0^2\right)\right] \big[\gamma _0^3 \left(g_R+g_S\right) \left(1 - \eta ^2 \left(1-T_L\right)\right)\\
&\quad +\gamma _0 \left(\eta ^2 \left(g_S+g_R \left(2-T_L\right)\right)+g_S\,T_L\right)+\eta\,\gamma_0^2 \left(g_R\,T_L+2 g_S\,T_L+g_R\right)-\gamma_0\,g_S\\
&\quad +\eta\,g_R \left(1-T_L\right)\big]\,\big[g_R \left(\gamma _0^2 \left(\eta\,\gamma _0+2\right)+\eta  \left(\eta +\gamma _0\right)+(1-\gamma _0^2) \left(1+\eta\,\gamma _0\right) T_L -1\right)\\
&\quad +g_S \left(T_{L}\,\left(1+\eta\,\gamma _0\right){}^2-\left(1-\eta^2\right) \left(1-\gamma _0^2\right)\right)\big]+\left(\eta +\gamma _0\right) \left(1+\eta\,\gamma_0\right) g_S \left(1-T_L\right)\\
&\quad\times\Big[\left(T_R-1\right) \left[\gamma _0 \left(\eta ^2 \left(g_S-g_R \left(T_L-2\right)\right)+g_S\, T_L\right)+\gamma _0^3 \left(g_R+g_S\right)\right.\\
&\quad\left.\times \left(1-\eta ^2 \left(1-T_L\right)\right)+\eta\,\gamma _0^2 \left(g_R\,T_L+2 g_S\,T_L+g_R\right) + \eta\,g_R \left(1-T_L\right)-\gamma_0\, g_S\right]{}^2\\
&\quad +\big[g_R \left( \left(2+\eta\,\gamma _0\right)\gamma_{0}^{2}+\eta  \left(\eta +\gamma _0\right)+(1-\gamma _0^2) \left(1+\eta\,\gamma _0\right) T_L-1\right)\\
&\quad +g_S \left(\left(1 + \eta\,\gamma _0\right){}^2 T_L-\left(1-\eta ^2\right) \left(1-\gamma _0^2\right)\right)\big]{}^2\Big].
\end{split}
\end{equation}
Extracting the leading term of $\gamma_{0}$ in $\eta$, we get $\gamma_{0}=A_{0}\,\eta^{1/\delta}$ for $\eta=0^{+}$, where $A_{0}\neq 0$ and $\delta>0$ are determined from $Q(\eta,\gamma_{0})= 0$. If we set $\delta=3$, we obtain for the leading contribution to $Q(\eta,\gamma_{0})$ the following equation
\begin{equation}
\begin{split}\label{Q}
Q(\eta,\gamma_{0})\sim&\left(1-T_L\right){}^2 \Big[g_R\,g_S\left(g_L+g_S\right)\left[3(1-T_L)+\left(2+T_L\right)A_{0}^3\right]+g_R^3 \left(1+T_L+A_{0}^3\right)\\
&\quad +g_S^3 \left[1-T_{L}+\left(T_{R}+T_{L}-T_{R}\,T_{L}\right)A^{3}_{0}\right]\Big]\,\eta = 0,
\end{split}
\end{equation}
from which we obtain the non-trivial value for $A_{0}$ (ie $A_{0}\neq 0$), which simultaneously satisfy the condition $\text{Im}[A_{0}]>0$ for $\eta=0^{+}$ due to $\nu (x)>0$. We get $A_{0}=\text{e}^{\ii\pi/3}\,\text{Abs}\left[A_{0}\right]$, where
\begin{equation}
\text{Abs}\left[A_{0}\right]=\frac{\left(g_{L}+g_{R}\right)(1-T_{L})^{1/3}}{\,\,\left[\,g_{R}^{3}+g_{R}\,g_{S}\left(g_{R}+g_{S}\right)\left(2+T_{L}\right)+g_{S}^{3}\left(T_{L}+T_{R}-T_{L}\,T_{R}\right)\,\right]^{1/3}\,\,}.
\end{equation}
Substituting $\gamma_{0}=A_{0}\,\eta^{1/3}$ in \eqref{eq_K}, we obtain the leading contribution of the pseudo-current as $K(\eta)\equiv K_{L}(\eta,\gamma_{0}) = -2\,g_{L}\,A_{0}\,\eta^{1/3}/T_{L}$. We denote the leading term of \eqref{nu} as $\nu(x)=-(1/\pi)\,\text{Im}\left[K_{L}(\eta,\gamma_{0})\right]=\mathcal{D}^{(0)}_{L}\,x^{1/3} $, where
\begin{equation}\label{nu_L}
\mathcal{D}^{(0)}_{L}=\frac{\sqrt{3}}{\pi}\,\frac{g_{L}\,\text{Abs}[A_{0}]}{T_{L}}>0,
\end{equation}
where we used $\eta=\tanh(x) \sim x$ for $x=0^{+}$. Finally, inserting $\gamma_{0}=A_{0}\,\eta^{1/3}$ into equation \eqref{gamma0p} for $\eta=0^{+}$, we get $\gamma_{0}^{\p} = A_{0}^{\p}\,\eta^{1/3}$, where $A_{0}^{\p}\equiv A_{0}\left[1-g_{L}\,g_{S}^{-1}/\left(1-T_{L}\right)\right]$.\\

\textbf{(ii)} Expanding the pseudo-currents $K_{L}(\eta,\bar{\xi})$ and $K_{R}(\bar{\xi}^{\,\p})$ defined in (\ref{KKL}) and (\ref{KKR}) in powers of $\epsilon$  through $\bar{\xi}=\gamma_{0} + \sum_{m\geqslant 1}\epsilon^{m}\,\gamma_{m}$ and $\bar{\xi}^{\p}=\gamma_{0}^{\p} + \sum_{m\geqslant 1}\epsilon^{m}\,\gamma_{m}^{\p}$, we obtain
\begin{subequations}\label{cel} 
\begin{align}
&K_{L}(\eta,\bar{\xi})= K_{L}(\eta,\gamma_{0})+2\,g_{L}\sum_{m\geqslant 1}\epsilon^{m}\,\left[\,\gamma_{m}\,E_{L}(\eta,\gamma_{0})+F^{(m)}_{L}(\eta,\vec{\gamma}_{[m]})\,\right],\quad\text{with}\quad F^{(1)}_{L}\equiv 0\label{KL_e}\\
&K_{R}(\bar{\xi}^{\,\p})= K_{R}(\gamma_{0}^{\p})+2\,g_{R}\sum_{m\geqslant 1}\epsilon^{m}\,\left[\,\gamma_{m}^{\,\p}\,E_{R}(\gamma_{0}^{\,\p})+ F^{(m)}_{R}(\vec{\gamma}^{\,\p}_{[m]})\,\right],\quad\text{with}\quad F^{(1)}_{R}\equiv 0\label{KR_e},
\end{align}
\end{subequations}
where $\vec{\gamma}_{[m]}\equiv(\gamma_{0}, \ldots, \gamma_{m-1})$ and $\vec{\gamma}^{\,\p}_{[m]}\equiv(\gamma_{0}^{\,\p}, \ldots, \gamma_{m-1}^{\,\p})$. Furthermore
\begin{subequations}
\begin{align}
&E_{L}(\eta,\gamma_{0}) =\frac{(1-\eta^2)\,\left[\,1-T_{L}+\eta^2 + 2\left(2-T_{L}\right)\eta\,\gamma_{0}+ \left(\,1+(1-T_{L})\eta^2\,\right)\,\gamma_{0}^2\,\right]}{\left[\,1-T_{L}-\eta^2-2\,T_{L}\eta\,\gamma_{0}-\left(\,1-(1-T_{L})\,\eta^2\,\right)\,\gamma^2_{0}\,\right]^{2}},\\
&E_{R}(\gamma^{\,\p}_{0}) =\frac{1+\left(1-T_{R}\right)\gamma_{0}^{\,\p\,2}}{\left[\,1-\left(1-T_{R}\right)\gamma_{0}^{\,\p\,2}\,\right]^{2}}.
\end{align}
\end{subequations}
Note also that $E_{L}(0,\gamma_{0})=A_{L}(-\ii\,\gamma_{0})$ and $E_{R}(\gamma^{\,\p}_{0})=A_{R}(-\ii\,\gamma^{\,\p}_{0})$. Similar to equation \eqref{BL}, we define
\begin{equation}\label{FL}
F^{(m)}_{L}(\eta,\vec{\xi}_{[m]})=\sum^{m}_{\substack{
   a_{1},\,a_{2},\,\ldots\, ,\,a_{m-1} = 0 \\
   \left(\sum_{j=1}^{m-1}j\,a_{j}=m\right)
  }}\gamma_{1}^{a_{1}}\,\gamma_{2}^{a_{2}}\,\cdots\,\gamma_{m-1}^{a_{m-1}}\,\,f_{L}^{\mathcal{P}}(\eta,\gamma_{0}),\quad\text{for}\quad m\geqslant 2,
\end{equation}
and also we define $F_{R}^{(m)}(\vec{\gamma}_{[m]}^{\,\p})= -\ii\,B^{(m)}_{R}(\ii\,\vec{\gamma}_{[m]}^{\,\p})$. For $u=\eta^{1/\delta}=0^{+}$ and $\delta>0$, we obtain the leading term of $F^{(m)}_{L}(\eta, u\,\vec{\gamma}_{[m]})$ and $F^{(m)}_{R}(u\,\vec{\gamma}_{[m]}^{\,\p})$ as
\begin{subequations}
\begin{align}
& F^{(m)}_{L}(\eta, u\,\vec{\gamma}_{[m]})=u^{3}\left[\,\Theta(1-\delta)\,\frac{\left(2+T_{L}\right)}{\left(1-T_{L}\right)^2}\sum^{m-1}_{\substack{
   i, j\,=\,1 \\
   \left(i + j\,=\,m\right)
  }} \gamma_{i}\,\gamma_{j}\,-\,\Theta(\delta-1)\,
  B^{(m)}_{L}(\vec{\gamma}_{[m]})\right] + \mathcal{O}(u^4),\\
& F^{(m)}_{R}(u\,\vec{\gamma}_{[m]}^{\,\p})= - u^3\,B^{(m)}_{R}(\vec{\gamma}^{\,\p}_{[m]}) + \mathcal{O}(u^4),
\end{align}
\end{subequations}
where we adopted for the Heaviside function $\Theta(0)\equiv 1$.

Substituting \eqref{IS_e} into \eqref{KS}, we obtain the expansion of $K_{S}(\bar{\chi})$ as
\begin{equation}\label{KSX}
K_{S}(\bar{\chi}) =-\ii\,\tilde{I}_{S}(\ii\,\bar{\chi}) = 2\,g_{S}\,\bar{\chi}_{0} + 2\,g_{S}\sum_{m\geqslant 1}\epsilon^{m}\left(\gamma_{m}\,\overline{X} + \gamma^{\,\p}_{m}\,\overline{Y} + \overline{W}^{(m)} \right),
\end{equation}
where $\overline{X}=\check{X}(\ii\,\gamma_{0},~\ii\,\gamma^{\,\p}_{0})$, $\overline{Y}=\check{Y}(\ii\,\gamma_{0},~\ii\,\gamma^{\,\p}_{0})$ and $\overline{W}^{(m)}=-\ii\,\check{W}^{(m)}(\ii\,\vec{\gamma}_{[m]},~\ii\,\vec{\gamma}_{[m]}^{\,\p})$.

Inserting \eqref{KL_e}, \eqref{KR_e} and \eqref{KSX} into \eqref{eq_K}  we get the following set of equations for the set $\lbrace\gamma_{m},\gamma^{\p}_{m}\rbrace$ 
\begin{subequations}
\begin{gather}
K_{L}(\eta,\gamma_{0})=2\,g_{S}\,\bar{\chi}_{0}=K_{R}(\gamma_{0}^{\,\p}),\quad\text{to zero order in}\;\epsilon\label{gamma_0}\\
g_{L}\left(\gamma_{m}\,E_{L}+F_{L}^{(m)}\right)= g_{S}\,\left(\gamma_{m}\,\overline{X}+\gamma^{\,\p}_{m}\,\overline{Y} + \overline{W}^{(m)}\right)=g_{R}\left(\gamma^{\,\p}_{m}\,E_{R} + F^{(m)}_{R}\right),\,\,\text{for}\,\,m\geqslant 1.\label{gamma_n}
\end{gather}
\end{subequations}

The solution of \eqref{gamma_0} is then given by $Q(\eta, \gamma_{0})=0$ in agreement with (\ref{eq_xi0}) and (\ref{Q}). Solving \eqref{gamma_n} for $(\gamma_{m},\gamma^{\p}_{m})$ with $m\geqslant 1$, we find
\begin{align}\label{A42}
\gamma_{m}=\frac{\overline{\mathcal{Y}}^{(m)}}{\overline{\mathcal{W}}^{(m)}}\qquad\text{and}\qquad\gamma_{m}^{\p}=\frac{\overline{\mathcal{X}}^{(m)}}{\overline{\mathcal{W}}^{(m)}}\, ,
\end{align}
where
\begin{subequations}\label{A43}
\begin{align}
&\overline{\mathcal{Y}}^{(m)} \equiv g_{R}\,E_{R}\,\big(g_{L}\,F^{(m)}_{L} -  g_{S}\,\overline{W}^{(m)}\big) + g_{S}\,\big(\,g_{R}\,F_{R}^{(m)} - g_{L}\,F_{L}^{(m)}\,\big) \overline{Y},\label{Y_over}\\
&\overline{\mathcal{X}}^{(m)} \equiv g_{L}\,E_{L}\big(\,g_{R}\,F^{(m)}_{R}-g_{S}\,\overline{W}^{(m)}\,\big) + g_{S}\,\big(\,g_{L}\,F_{L}^{(m)} - g_{R}\,F_{R}^{(m)}\,\big)\,\overline{X},\label{X_over}\\ 
&\overline{\mathcal{W}}^{(m)} \equiv g_{S}\,\big(\,g_{L}\,E_{L}\,\overline{Y} + g_{R}\,E_{R}\,\overline{X}\,\big)-g_{L}\,g_{R}\,E_{L}\,E_{R}.\label{W_over}
\end{align}
\end{subequations}

The solution of \eqref{gamma_0} for $\gamma_{0}$ and $\gamma_{0}^{\,\p}$ was obtained in part \textbf{(i)} of \ref{Appen_A3} as $\gamma_{0}=A_{0}\,\eta^{1/3}$ and $\gamma_{0}^{\p}=A_{0}^{\p}\,\eta^{1/3}$ with $\eta=0^{+}$ and $A_{0}$ and $A_{0}^{\p}$ finite. Thus, for $\gamma_{a}=\bar{y}_{a}\,u\,$ and $\,\gamma_{a}^{\,\p}=\bar{x}_{a}\,u$  (for $0\leqslant a \leqslant m-1$)  with $u=\eta^{1/3}=0^{+}$  and a finite set $\lbrace \bar{y}_{a}, \bar{x}_{a}\rbrace$, we can express $\overline{\mathcal{Y}}^{(m)}\propto u^{3}$, $\overline{\mathcal{X}}^{(m)}\propto u^{3}$ and $\overline{\mathcal{W}}^{(m)}\propto\eta^{2/3}=u^{2}$. By induction, we find $\gamma_{m} = u\,\bar{y}_{m}$ and $\gamma^{\,\p}_{m} = u\,\bar{x}_{m}$ with $m\geqslant 1$ for $u=0^{+}$ and $\bar{y}_{m}$ and $\bar{x}_{m}$ finite. Therefore,  $\bar{\xi}=\bar{y}\,\eta^{1/3}$ and $\bar{\xi}^{\,\p}=\bar{x}\,\eta^{1/3}$, where $\bar{y}=A_{0}+\sum_{m\geqslant 1}\epsilon^{m}\,\bar{y}_{m}$ and $\bar{x}=A^{\p}_{0}+\sum_{m\geqslant 1}\epsilon^{m}\,\bar{x}_{m}$ for $\epsilon>0$, and inclusive when $\epsilon=1$, where $\bar{y}$ and $\bar{x}$ are finite by construction. Finnaly, 
\begin{subequations}
\begin{equation}\label{nuxy}
\nu(x)=-\frac{1}{\pi}\,\text{Im}\left[K_{L}(\eta,\bar{\xi})\right]=\mathcal{D}_{L}\,x^{1/3},
\end{equation}
where we defined the non-universal amplitude
\begin{equation}\label{DL}
\mathcal{D}_{L}= \mathcal{D}_{L}^{(0)} + \sum_{m\geqslant 1}\epsilon^{m}\,\mathcal{D}_{L}^{(m)} \quad\text{with}\quad\mathcal{D}_{L}^{(m)}=-\frac{2}{\pi}\frac{g_{R}\,g_{S}\left(1-T_{L}\right)}{\left(g_{R}+g_{S}\right)\,T_{L}}\,\text{Im}\left[\bar{y}_{m}\right].
\end{equation}
\end{subequations}
The sum represented in \eqref{DL} has guaranteed convergence for $\epsilon=1$ by virtue of the convergence of $\bar{y}\vert_{\epsilon=1}=A_{0}+\sum_{m\geqslant 1}\bar{y}_{m}$.

The behavior of $\nu(x)$ at $\zeta_{R}=\zeta_{c}$ can be obtained through the exchange $L\leftrightarrow R$. This is justified from the change of variables $\theta\to\phi-\theta^{\p}$ and $\theta^{\p}\to\phi-\theta$ in \eqref{eq_K} and the set of equations \eqref{LRS}. We thus may write $\nu(x) = \mathcal{D}\, x^{1/3}$ for $\zeta=\zeta_{c}$, where $\mathcal{D}=\mathcal{D}_{R}\,\Theta(g_{L}-g_{R})+\mathcal{D}_{L}\,\Theta(g_{R}-g_{L})$ (for $\epsilon=1$), which completes the proof of Corollary \ref{coro_2}.

\subsection{Corollary 3}\label{Appen_A6}

Consider the conservation of pseudo-current shown in equations \eqref{eq_K} and \eqref{LRS}, with the replacement $R(\bar{\chi})\to\epsilon\,R(\bar{\chi})$, and for $\zeta=\zeta_{c}\,(1-t)$ via \eqref{zetaL} with $t=0^{-}$. We expand the variables $\bar{\xi}$ and $\bar{\xi}^{\p}$ as $\bar{\xi}=\gamma_{0} + \sum_{m\geqslant 1}\epsilon^{m}\,\gamma_{m}$ and $\bar{\xi}^{\p}=\gamma_{0}^{\,\p} + \sum_{m\geqslant 1}\epsilon^{m}\,\gamma_{m}^{\,\p}$. Taking into account \eqref{cel} - \eqref{KSX}, we obtain the set of equations \eqref{A41ab} for the determination of the set $\lbrace\gamma_{m},\gamma_{m}^{\,\p}\rbrace$ for $m\geqslant 0$.

The solution of \eqref{xi_0} for the variable $\gamma_{0}^{\,\p}$ is expressed in terms of $\gamma_{0}$ in equation \eqref{gamma0p}, whereas the variable $\gamma_{0}$ is determined from $Q(\eta,\gamma_{0})=0$ using \eqref{eq_xi0}.
Setting $\gamma_{0}=\bar{y}_{0}\,u$  and $\gamma_{0}^{\,\p}=\bar{x}_{0}\,u$ with $u=(-t)^{\beta^{\p}}=0^{+}$ and $\eta=h\,(-t)^{\Phi/2}=0^{+}$ for $h$ and $\lbrace\bar{y}_{a}\rbrace$ finite, we extract the leading term of $\gamma_{0}$ in $-t$ for the polynomial $Q(\eta,\gamma_{0})$ for  $0<\beta^{\p}<1$ and $\Phi>0$. We get
\begin{equation}\label{Qt}
0=Q(\eta,\gamma_{0}) \sim \left\{
  \begin{array}{lr}
    V\,h\,(-t)^{\Phi/2}, &\text{for}\quad 1/2<\beta^{\p}<1\;\;\text{and}\;\;\Phi>0;\\
    \left[\,S\,\bar{y}_{0} + U\,\bar{y}_{0}^{3}\,\delta_{\beta^{\p},1/2}\,\right]\,(-t)^{1+\beta^{\p}}, &\text{for}\;0<\beta^{\p}\leqslant 1/2\;\;\text{and}\;\;\Phi > 3;\\
    Q_{0}(\beta,\Phi), &\text{for}\; 0<\beta^{\p}<1/2\;\;\text{and}\;\;0<\Phi<3;\\
    V\,h\,(-t)^{\Phi/2}, &\text{for}\;\; \beta^{\p}=1/2\;\;\text{and}\;\;0<\Phi<3;\\
   \left[\,\left(V\,h + S\,\bar{y}_{0}\right)\delta_{\beta^{\p},1/2} + U\,\bar{y}_{0}^{3}\,\right]\,(-t)^{\beta^{\p}\,\Phi}, &\text{for}\;\;0<\beta^{\p}\leqslant 1/2\;\;\text{and}\;\;\Phi=3.
      \end{array}
\right.
\end{equation}
where $Q_{0}(\beta^{\p},\Phi)\equiv V\,h\,(-t)^{\Phi/2}\,\Theta(6\,\beta^{\p}-\Phi-0^{+}) + \left[V\,h\,\delta_{\,6\beta^{\p},\Phi} + U\,\bar{y}_{0}^{3}\,\Theta(\Phi-6\,\beta^{\p})\right]\,(-t)^{2\beta^{\p}\Phi}$ and we also defined
\begin{subequations}
\begin{align}
V &= g_{R}\,g_{S}^2\,(1-T_{L})^{3},\\
S &= - g_{R}\,g_{S}^{2}\,(1-T_{L})^2\,(g_R +g_S\,T_{L})\,(g_{R}+g_{S})^{-1},\\
U &= \frac{g_{R}\,g_{S}^{2}\,(1-T_{L})^{2}}{(g_{R}+g_{S})^{3}}\,\left[\,g_{R}^{3} + g_{R}\,g_{S}\,(g_{R}+g_{S})\,(2+T_{L}) + g_{S}^{3}\,(T_{L}+T_{R}-T_{L}\,T_{R})\,\right].
\end{align}
\end{subequations}

From \eqref{Qt}, we see that only in the domain $\Phi=6\,\beta^{\p}$ for $0<\beta^{\p}\leqslant 1/2$ or $0<\Phi\leqslant 3$, we obtain a non-trivial condition for $h$ or $\bar{y}_{0}$ in the form of 
\begin{equation}\label{A47}
0=Q(\eta,\gamma_{0}) \sim \left\{
  \begin{array}{lr}
\left(V\,h + U\,\bar{y}_{0}^{3}\right)(-t)^{12\,\beta^{\p\,2}}, & \text{for}\quad  0<\beta^{\p}<1/2;\\
\left(V\,h + S\,\bar{y}_{0} + U\,\bar{y}_{0}^{3}\right)\,(-t)^{3/2}, & \text{for}\quad \beta^{\p}=1/2.
  \end{array}
\right.
\end{equation}

The solution of \eqref{gamma_n} is determined by \eqref{A42} and \eqref{A43} for the set $\lbrace\gamma_{m},\gamma_{m}^{\,\p}\rbrace$ with $m\geqslant 1$. If we assume that $\gamma_{a}=\bar{y}_{a}\,u$ and $\gamma_{a}^{\,\p}=\bar{x}_{a}\,u$ (for $0\leqslant a\leqslant m-1$) with $u=(-t)^{\beta^{\p}}=0^{+}$ and $\lbrace\bar{y}_{a},\bar{x}_{a}\rbrace$ finite, we see that $\overline{\mathcal{Y}}^{(m)}\propto u^{3}$, $\overline{\mathcal{X}}^{(m)}\propto u^{3}$ and $\overline{\mathcal{W}}^{(m)}\propto (-t) =u^{1/\beta^{\p}}$. Therefore by induction, from \eqref{A42} we obtain $\gamma_{m}\propto u^{3-1/\beta^{\p}}= u$ for $\beta^{\p}=1/2$ $\;\forall\,m$ (and similarly for $\gamma_{m}^{\,\p}$). Then according to \eqref{A47}, we adjust the exponent $\Phi=6\,\beta^{\p}=3$. 

Thus, we obtain $\bar{\xi}=\bar{y}\,(-t)^{\beta^{\p}}$ and $\bar{\xi}^{\,\p}=\bar{x}\,(-t)^{\beta^{\p}}$ where $\bar{y}=\bar{y}_{0}+\sum_{m\geqslant 1}\epsilon^{m}\,\bar{y}_{m}$ and $\bar{x}=\bar{x}^{\p}_{0}+\sum_{m\geqslant 1}\epsilon^{m}\,\bar{x}_{m}$ for  $\epsilon>0$. We may thus write
\begin{subequations}
\begin{equation}
\nu(x)=-\frac{1}{\pi}\,\text{Im}\left[K_{L}(\eta,\bar{\xi})\right]=\mathcal{B}^{\,\p}_{L}\,(-t)^{1/2}\,\Theta(-t),
\end{equation}
where we defined the non-universal amplitude
\begin{equation}\label{BLp}
\mathcal{B}_{L}^{\,\p}= \mathcal{B}_{L}^{\,\p(0)} + \sum_{m\geqslant 1}\epsilon^{m}\,\mathcal{B}_{L}^{\,\p(m)} \quad\text{with}\quad\mathcal{B}_{L}^{\,\p(m)}=-\frac{2}{\pi}\frac{g_{R}\,g_{S}}{\left(g_{R}+g_{S}\right)}\,\text{Im}\left[\bar{y}_{m}\right].
\end{equation}
\end{subequations}
The sum represented in \eqref{BLp} has guaranteed convergence for $\epsilon=1$ by virtue of the convergence of $\bar{y}\vert_{\epsilon=1}=\bar{y}_{0}+\sum_{m\geqslant 1}\bar{y}_{m}$ by construction.

The behavior of $\nu(x)$ at $\zeta_{R}=\zeta_{c} + 0^{+}$ can be obtained through the exchange $L\leftrightarrow R$. This is justified from the change of variables $\theta\to\phi-\theta^{\p}$ and $\theta^{\p}\to\phi-\theta$ in equation \eqref{eq_K} and the set \eqref{LRS}. We thus may write $\nu(x) = \mathcal{B}^{\,\p}\, (-t)^{1/2}$ for $\zeta=\zeta_{c} + 0^{+}$, where $\mathcal{B}^{\,\p}=\mathcal{B}_{R}^{\,\p}\,\Theta(g_{L}-g_{R})+\mathcal{B}_{L}^{\,\p}\,\Theta(g_{R}-g_{L})$ (for $\epsilon=1$), which completes the proof of Corollary \ref{coro_3}.

\subsection{Theorem 2}\label{Appen_A4}

The pseudo-current of the sample, $I_{S}(\varphi)$, is an analytical function in the variable $1/s$ and therefore we can expand it in power series as follows
\begin{equation}\label{Laurent}
I_{S}(\varphi)=\frac{N_{S}}{s}\,\varphi + \mathcal{O}(1/s^2),
\end{equation}
which clearly satisfies the conditions $I_{S}(0)=0$ and $I_{S}^{\p}(0)= N_{S}/s\equiv g_{S}$. This expansion guarantees the leading term $I_{S}(\varphi)=g_{S}\,\varphi$ expected for $s\gg 1$. In order to evaluate the double-scaling limit, defined by taking the diffusive limit ($s\gg 1$) and simultaneously setting the condition $\zeta=\zeta_{c}/\lambda$, we calculate the pseudo-potentials $\theta$ and $\theta^{\p}$ in power series of $1/s$ in the pseudo-current conservation law, equation \eqref{kirchhoff}, by determining the coefficients $\lbrace l_{n}, r_{n}\rbrace$, defined from $\theta=l_{0} + \sum_{n\geqslant 1} l_{n}\,s^{-n}$ and $\theta^{\p}=r_{0} + \sum_{n\geqslant 1} r_{n}\,s^{-n}$. 
The condition $\zeta=\zeta_{c}/\lambda$ is considered separately in two intervals: \textbf{i)} $\lambda>1$ or $0<\zeta<\zeta_{c}$  and \textbf{ii)} $0<\lambda\leqslant 1$ or $\zeta\geqslant\zeta_{c}$.Note that if we choose $g_{R}>g_{L}$, the condition $\zeta=\zeta_{c}/\lambda$ yields $0<T_{R}\leqslant 1$ and $T_{L}\equiv T_{c}(s,\lambda)$, where
\begin{equation}
T_{c}(s,\lambda)=\frac{N_{S}\,N_{R}\,T_{R}\,\lambda }{N_{L}\,N_{R}\,T_{R}\,s + N_{S}\left(N_{L}\lambda + N_{R}\,T_{R}\right)}.
\end{equation}
\\
\textbf{(i)} 
The condition $0<\zeta<\zeta_{c}$ is implemented by the representation $T_{L}=(1-\alpha)\,T_{o}\,+\,\alpha\,T_{c}(s,1)$, where $T_{o}\equiv N_{R} T_{R} / N_{L}$ and $0<\alpha<1$. From \eqref{kirchhoff}, we determine the first coefficients of $\theta$ and $\theta^{\p}$ by expanding in power series of $1/s$ the pseudo-currents $I(\phi-\theta;T_{L})$, $I(\theta^{\p};T_{R})$ and $I_{S}(\theta-\theta^{\p})$ (using \eqref{Laurent}). We find
\begin{subequations}
\begin{align}
&I(\phi-\theta; T_{L})\equiv I_{L}^{(0)}(\phi-l_{0})+I^{(1)}_{L}(\phi-l_{0},l_{1})\,\frac{1}{s}+\mathcal{O}(1/s^2),\label{IL}\\
&I_{S}(\theta-\theta^{\p})\equiv I^{(1)}_{S}(l_0 - r_0)\,\frac{1}{s} + \mathcal{O}(1/s^2),\label{IS}\\
&I(\theta^{\p};T_{R})\equiv I^{(0)}(r_{0})+I^{(1)}(r_{0},r_{1})\,\frac{1}{s}+\mathcal{O}(1/s^2).\label{IR}
\end{align}
\end{subequations}
From \eqref{kirchhoff}, we establish that $I^{(0)}_{L}(\phi-l_{0})=0=I^{(0)}_{R}(r_{0})$, therefore $l_{0}=\phi$ and $r_{0}=0$, where we used the property $I_{L,R}^{(0)}(0)=0$.  Next we obtain $I^{(1)}_{L}(0,l_{1})=I_{S}^{(1)}(\phi)=I^{(1)}_{R}(0,r_{1})$, which can be expressed explicitly as $N_{R}\,T_{R}\,(-1+\alpha)\,l_{1} = N_{S}\,\phi = N_{R}\,T_{R}\,r_{1}$. We have thus shown that the pseudo-current of the sample is  indeed given by $I(\phi)=g_{S}\,\phi + \mathcal{O}(1/s^2)$.\\

\textbf{(ii)} The condition $\zeta\geqslant\zeta_{c}$ is now represented by  $\,T_{L}=T_{c}(s,\lambda)$ (for $0<\lambda\leqslant 1$). Similarly to the previous case, from \eqref{kirchhoff} we determine the first coefficients $\lbrace l_{n}, r_{n}\rbrace$ of the expansion of the pseudo-potentials $\theta$ and $\theta^{\p}$ using the series in powers of $1/s$ of the pseudo-current $I(\phi-\theta;T_{L})$. We find
\begin{equation}
I(\phi-\theta;T_{L})=N_{S}\,\lambda\,\sin(\phi-l_{0})\,\frac{1}{s}+\mathcal{O}(1/s^{2})
\end{equation}
The expansion in series of powers of $1/s$ of the pseudo-currents $I_{S}(\theta-\theta^{\p})$ and $I(\theta^{\p},T_{R})$ for this case is the same as equations \eqref{IS}  and \eqref{IR}, respectively. Similarly to the previous case, the conservation of pseudo-current, equation \eqref{kirchhoff}, in order $1/s^{0}$ is $I^{(0)}(r_{0})=0$, thus $r_{0}=0$. Then, using equation  \eqref{kirchhoff} in order $1/s$ we get $N_{S}\,\lambda\,\sin(\phi-l_{0}) = N_{S}\,l_{0} = N_{R}\,T_{R}\,r_{1}$, where $l_{0}\equiv\phi-\bar{\phi}$ and $\phi=\bar{\phi}+\lambda\sin\bar{\phi}$. We have thus represented the pseudo-current of the system as $I(\phi)=g_{S}\,(\phi-\bar{\phi})+\mathcal{O}(1/s^{2})$.\\

Finally, note that we can rewrite the pseudo-current of system for arbitrary $\lambda$ as
$I_{\lambda}(\phi)=g_{S}(\phi-\bar{\phi}_{\lambda})$, where $\bar{\phi}_{\lambda}=\bar{\phi}\,\Theta(1-\lambda) + 0^{+}$ and $\phi-\bar{\phi}=\lambda\sin\bar{\phi}$, which concludes the proof.
\\

\subsection{Theorem 3}\label{Appen_A5}

Let $I_{0}$ denote the pseudo-current of the auxiliary sample described in Theorem 3. Let us now connect this auxiliary sample to an arbitrary anomalous diffusive conductor with dimensionless resistance $r$, where $r\equiv 1/g=s/N_{\text{ch}}$. Thus, the law of conservation of pseudo-current in the system is $I(\phi ;r)\equiv I_{0}(\phi-\theta)=I_{\lambda}(\theta)$, where $I_{\lambda}(\theta)$ is the pseudo-current of the anomalous conductor, which according to Theorem 2 is given by $I_{\lambda}(\theta)=(\theta-\bar{\theta}_{\lambda})/r$, where $\bar{\theta}_{\lambda}=\bar{\theta}\,\Theta(1-\lambda) + 0^{+}$ and $\theta-\bar{\theta}=\lambda\sin\bar{\theta}$. Therefore, we obtain $\theta=\bar{\theta}_{\lambda} + r I(\phi;r)$, where $\sin\bar{\theta}_{\lambda}=(r/\lambda)\,I(\phi; r)\,\Theta(1-\lambda) + 0^{+}$. Substituting the pseudo-potential $\theta$ into $I_{0}(\phi-\theta)$ and using the conservation law, we obtain the equation \eqref{sol_implicit}.
The above equation is an implicit solution to an equation for the pseudo-current $I(\phi;r)$ with a initial condition $I(\phi;0)=I_{0}(\phi)$.
Let us now show that $I(\phi,r)$ is also a solution of a non-linear partial differential equation. Performing the derivative with respect to $\phi$ and $r$ in \eqref{sol_implicit}, we obtain $\partial_{\phi}I=\left(1-\partial_{\phi}\bar{\theta}_{\lambda}-r\,\partial_{\phi}I\right)\,I^{\p}_{0}(\alpha)$ and $\partial_{r}I=-\left(\partial_{r}\bar{\theta}_{\lambda}+I+r\,\partial_{r}I\right)I^{\p}_{0}(\alpha)$, where $\alpha=\phi-\bar{\theta}_{\lambda} - r\,I$ and $I\equiv I(\phi;r)$.
Combining these equations so as to eliminate $I^{\p}_{0}(\alpha)$, we obtain $\partial_{r} I + I\,\partial_{\phi}I = \partial_{r}I\,\partial_{\phi}\bar{\theta}_{\lambda}-\partial_{\phi}I\,\partial_{r}\bar{\theta}_{\lambda}$. 
Calculating $\partial_{\phi}\bar{\theta}_{\lambda}$ and $\partial_{s}\bar{\theta}_{\lambda}$ from \eqref{sol_implicit}, we get $\partial_{\phi}\bar{\theta}_{\lambda}=\lambda^{-1}\,\Theta(1-\lambda)\,r\,\sec(\bar{\theta}_{\lambda})\,\partial_{\phi}I$ and $\partial_{r}\bar{\theta}_{\lambda}=\lambda^{-1}\,\Theta(1-\lambda)\,\sec(\bar{\theta}_{\lambda})\,\left(I + r\,\partial_{r}I\right)$. Finally, we may write
\begin{equation}
\frac{\partial I}{\partial r}+I\,\frac{\partial I}{\partial \phi}
=-\lambda^{-1}\,\Theta(1-\lambda)\,\sec\bar{\theta}\,I\,\frac{\partial I}{\partial\phi}.
\end{equation}
From the relation $\lambda\sin\bar{\theta}=r\,I$, we obtain $\lambda\cos\bar{\theta}=\sqrt{\lambda^{2}-\left(r\,I\right)^{2}\,}$. We have thus shown that \eqref{scale_dif} is the scaling equation of an anomalous diffusive conductor.

\section{Formulas}

The polynomials shown in \eqref{q_NS_eq} are given by
\begin{subequations}
\begin{align}
P_{2}(\omega) =& \,\frac{9}{8}\,(11\,\omega^3 + 45 \,\omega^2 + 117\,\omega +115),\label{P2}\\
P_{3}(\omega) =& \,\frac{9}{160}\,\bigl(\,3520\,\omega^{11} + 49600\,\omega ^{10} + 293791\,\omega^9 + 978549\,\omega^8 + 1823226\,\omega^7 + 1212486\,\omega^6 \nonumber\\
&\quad\quad - 1919328\,\omega^5- 3694920\,\omega^4 + 550038\,\omega^3 + 5464106\,\omega^2 + 4392305\,\omega\nonumber\\
&\quad\quad + 1297571\,\bigl),\label{P3}\\
P_{4}(\omega) =& \frac{27}{4480}\,\bigl(\,315392\,\omega^{19}+7598080\,\omega^{18}+79325568\,\omega^{17} + 474085248\,\omega^{16} + 1756668801\,\omega ^{15}\nonumber\\
&\quad\quad  + 3787621263\,\omega^{14} + 2447558517\,\omega^{13} - 10455279861\,\omega^{12} - 29457644463\,\omega^{11}\nonumber\\
&\quad\quad  - 11863257121\,\omega^{10} + 76868542861\,\omega^9 + 146026782099\,\omega^8 + 28698240027\,\omega^7\nonumber\\
&\quad\quad - 216239103147\,\omega^6 - 272699628345\,\omega^5 - 43327965063\,\omega^4 + 180716858643\,\omega^3\nonumber\\
&\quad\quad + 186296913981\,\omega^2 + 79839412151\,\omega + 13777134697 \,\bigl)\label{P4}.
\end{align}
\end{subequations}

\end{document}